\documentclass[aps,pra,floatfix,superscriptaddress,showpacs,twocolumn,nofootinbib,longbibliography]{revtex4-2}

\usepackage[english]{babel}
\usepackage[utf8]{inputenc}
\usepackage[T1]{fontenc}
\usepackage[most]{tcolorbox}
\usepackage{enumitem}       

\usepackage{amsmath,amssymb,amsfonts,mathtools,dsfont}
\usepackage{bm, svg, esvect}  
\usepackage{bbm}   
\usepackage{braket}
\usepackage{physics} 
\usepackage{amsthm, dsfont} 
\usepackage{soul} 
\usepackage{orcidlink}

\usepackage{etoolbox}
\apptocmd{\normalsize}{%
  \setlength{\abovedisplayskip}{6pt plus 2pt minus 2pt}%
  \setlength{\belowdisplayskip}{6pt plus 2pt minus 2pt}%
  \setlength{\abovedisplayshortskip}{0pt plus 2pt}%
  \setlength{\belowdisplayshortskip}{4pt plus 2pt minus 2pt}%
}{}{}

\usepackage{graphicx}
\usepackage{booktabs}
\usepackage{placeins}   

\usepackage[dvipsnames]{xcolor}
\usepackage{enumitem}
\usepackage{csquotes}
\usepackage{tikz}
\hypersetup{
    hidelinks,
    colorlinks   = true,
    citecolor    = RoyalBlue,
    linkcolor    = RoyalBlue,
    urlcolor     = RoyalBlue
}

\DeclareMathOperator{\Ei}{Ei}

\begin{document}
\preprint{APS/123-QED}
\title{Parameter Estimation in a Continuously Monitored Non-Markovian Quantum System}
\author{Erik L. André$^{\orcidlink{0009-0008-8615-8058}}$}
\email{erik.andre@tuwien.ac.at}
\affiliation{Technische Universität Wien, Atominstitut, Vienna Center for Quantum Science and Technology, Stadionallee 2, 1020 Vienna, Austria}
\affiliation{Institute for Quantum Optics and Quantum Information (IQOQI), Austrian Academy of Sciences, Boltzmanngasse 3, 1090 Vienna, Austria}

\author{Pharnam Bakhshinezhad$^{\orcidlink{0000-0002-0088-0672}}$}
\affiliation{Technische Universität Wien, Atominstitut, Vienna Center for Quantum Science and Technology, Stadionallee 2, 1020 Vienna, Austria}

\author{Patrick P. Potts$^{\orcidlink{0000-0001-6036-7291}} $}
\affiliation{Department of Physics and Swiss Nanoscience Institute, University of Basel, Klingelbergstrasse 82, 4056 Basel, Switzerland}

\author{Luis A. Correa$^{\orcidlink{0000-0002-7357-1328}}$}
\affiliation{Instituto~Universitario~de~Estudios~Avanzados~(IUdEA),~Universidad~de~La~Laguna,~La~Laguna~38203,~Spain}
\affiliation{Secci\'on de F\'isica, Facultad de Ciencias, Universidad de La Laguna, La Laguna 38203, Spain}

\author{Mohammad Mehboudi$^{\orcidlink{0000-0002-0398-9200}}$}
\email{mohammad.mehboudi@gmail.com}
\affiliation{Technische Universität Wien, Atominstitut, Vienna Center for Quantum Science and Technology, Stadionallee 2, 1020 Vienna, Austria}

\date{\today}
\begin{abstract}
    Continuous monitoring is a powerful tool for analyzing quantum systems and is increasingly used as a non-demolition technique in quantum metrology. In this context, noisy data acquired through continuous monitoring can be used to precisely pinpoint unknown parameters of the system. However, extending the theoretical framework that connects these data to the underlying parameters beyond Markovian dynamics is notoriously difficult, primarily because such dynamics cannot be expressed as completely positive divisible maps. To overcome this, we propose a method based on the reaction coordinate mapping to extend these parameter-estimation techniques beyond the Markovian regime. Our approach specifically targets linear systems with \textit{non-Markovian} dynamics undergoing Gaussian continuous measurements, such as homodyne detection. 
    Within this framework, we analyze Bayesian estimation and provide an analytical expression for the Fisher information, alongside the asymptotic scaling of the estimation precision for arbitrary parameters.
    Finally, we demonstrate the efficacy of our method through the example of thermometry of a bosonic bath. 
\end{abstract}
\maketitle
\noindent{\it Introduction---}The 
theory of quantum metrology, or parameter estimation, is dedicated to designing measurement schemes for the inference of parameters of quantum systems~\cite{Tth2014,PARIS2009}. This requires preparing suitable probes, modeling information acquisition via probe--sample interaction, and analyzing the measured data to estimate unknown parameters of the system. Quantum metrology often aims at finding the fundamental bounds on estimation precision~\cite{PhysRevLett.96.010401,Giovannetti2011}, but an alternative and equally important application is the construction of practical models for experimental setups, accounting for noise~\cite{Escher2011,DemkowiczDobrzaski2012}, measurement imperfections~\cite{Len2022}, and dissipation~\cite{PhysRevLett.112.120405, Landi2025RMPNoneqBoundaryDrivenQS}. This is particularly relevant to optical interferometric setups, where realistic descriptions must account for effects such as phase diffusion, loss, and imperfect interferometric visibility~\cite{DemkowiczDobrzaski2015}. Such models characterize both the probe dynamics and the measurement process, while providing a meaningful way of interpreting the collected data.

In particular, the tools of quantum parameter estimation have been used to characterize continuous monitoring~\cite{Jacobs2006,PhysRevA.87.032115,PhysRevLett.132.050801}, providing new insights as to how to enhance noisy quantum metrology by harvesting information locked away in the environment~\cite{Albarelli2018,PhysRevLett.125.200505}. Continuous monitoring can be advantageous compared to prepare--measure--reset protocols, since it naturally accounts for the total protocol time as a resource, without any hidden preparation or reset times~\cite{PhysRevA.111.L020401}. Furthermore, continuous monitoring is the standard measurement approach in various platforms, including superconducting circuits~\cite{Vijay2012} and collective spin systems~\cite{Kong2020}, as well as cavity-QED, optomechanical and optical platforms~\cite{Zhang2017,khan2021efficient}. 

\begin{figure}
    \includegraphics[width=1\columnwidth]{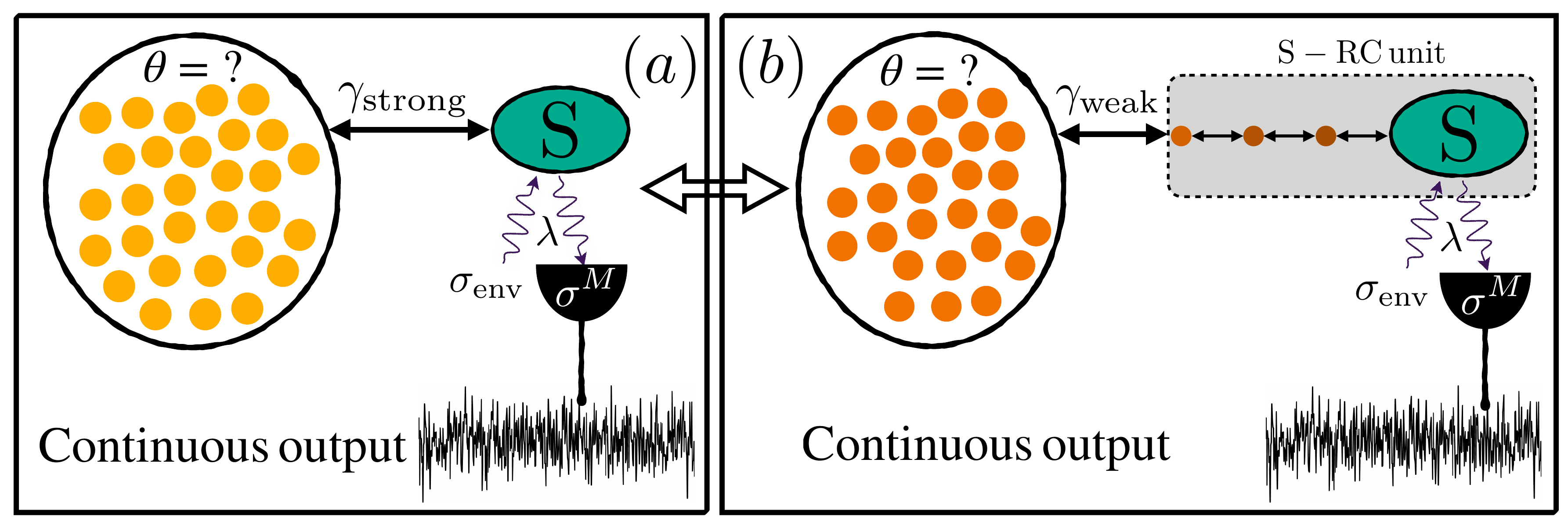}
    \caption{(a) We consider a continuously monitored system that is strongly coupled to a bath and therefore exhibits non-Markovian dynamics. The output signal is not straightforwardly interpretable; hence it cannot be immediately used to infer unknown bath parameters. (b) The reaction coordinate mapping resolves this issue by recasting the problem as a Markovian one, allowing the output signal to be analyzed efficiently. Our proposal works for arbitrary Gaussian measurements, represented by the covariance matrix $\sigma^M$ (e.g., homodyne detection), 
    performed on an environmental mode initially prepared in a Gaussian state represented by its covariance matrix $\sigma_{\rm env}$.}
    \label{fig:RC}
\end{figure}

The data generated by continuous monitoring can be used to infer unknown parameters of a sample by resorting to either the frequentist~\cite{PhysRevLett.112.170401,Kiilerich2014,PRXQuantum.5.020201,yxqg-93jz,Radaelli2026parameterestimation} or, as we do here, the Bayesian~\cite{PhysRevA.94.032103,molenda2026unlocking,Boeyens2023,k7nk-lrwd,amoros2025tracking} paradigm. Specifically, we focus on linear Gaussian quantum systems subject to Gaussian continuous measurements, such as homodyne detection, which is standard in quantum-optical setups. Crucially, the customary analysis of a continuous noisy signal relies on sequential updates of the conditional state and the associated likelihood, and demands that the underlying dynamics be completely positive divisible (CP-divisible), or \textit{Markovian} \cite{Wiseman2008NononMarkovianQuantumTrajectory}. Extending the quantum metrology of continuous signals into the non-Markovian (i.e., CP-indivisible) regime\footnote{Here, we equate `Markovianity' to `CP-divisibility'; for alternative notions of Markovianity and their potential equivalence with CP-indivisibility see, e.g., Refs.~\cite{Breuer2016NonMarkvianityReview,PhysRevLett.123.040401,PhysRevLett.118.120501}.} remains a major open challenge. This is precisely the problem that we tackle in this Letter.  
\newline

\noindent{\it Bypassing CP-indivisibility---}In order to overcome the CP-indivisibility issue, we resort to the reaction coordinate (RC) mapping, a well-established Markovian embedding technique. Originally proposed in chemical physics to treat friction in electron-transfer dynamics~\cite{GargOnuchicAmbegaokar1985}, the method has since been widely used to describe open quantum systems in highly structured environments~\cite{Iles_PRA2014,Nazir2018ReactionCoordinate, Strasberg_2016, correa_pushing_2019}, including current fluctuations~\cite{Mahadeviya2026}.
	
Let us first summarize the intuition behind the RC technique. As illustrated in Fig.~\ref{fig:RC}, a system in a structured linear bath can be equivalently thought of as being coupled---potentially strongly---to one or more collective degrees of freedom drawn from the environment. Such collective modes are referred to as reaction coordinates~\cite{Nazir2018ReactionCoordinate,Iles_PRA2014} and are, in turn, coupled to the residual bath, made up of the remaining environmental degrees of freedom. The reaction coordinates thus mediate dissipative energy exchanges between the system and the residual bath. The key observation is that whenever enough RCs have been singled out from the environment, the residual bath becomes unstructured \cite{MartinazzoVacchiniHughesBurghardt2011}, which translates into short-lived bath correlation functions. Furthermore, one can always find parameter regimes in which the mapping yields a weak coupling to the residual bath, so that the system--reaction coordinate (S--RC) unit may be described by a Gorini--Kossakowski--Lindblad--Sudarshan (GKLS) master equation \cite{lindblad1976generators,gorini1976completely}. Consequently, the S--RC unit would undergo a CP-divisible dynamics even if the system alone experiences strongly non-Markovian dissipation. If the system were being continuously monitored, its evolution could be cast as~\cite{wiseman_milburn_2009,jacobs_2014,PhysRevLett.129.050401} 
\begin{equation}\label{eq:cont_mon_divisible_evolution}
		\rho(t|\vv{W}_m) = \mathcal{M}_{W_m}(\lambda \delta t) \circ {\cal E}_{\delta t} \cdots \mathcal{M}_{W_1}(\lambda \delta t) \circ {\cal E}_{\delta t} \, \rho(0),
\end{equation}
where $\rho(0)$ is the initial state of the augmented S--RC unit, $\vv W_m = \{W_m, \dots, W_1\}$ is the sequence of observed outcomes of measurements performed on S, and $\delta t = t/m$, so that the limit $\delta t \to 0$ gives a continuous signal. Here, the superoperator ${\cal E}_{\delta t}$ is a CPTP map representing an infinitesimal evolution of the S--RC unit while undisturbed by the measurement. In turn, the maps $\mathcal{M}_{W_k}(\lambda \delta t)$ take an input of the S--RC unit into an output state conditioned on observing the outcome $W_k$, with $\lambda$ representing the strength of the measurement\footnote{In what follows, we take $\mathcal{M}_{W_k}(\lambda \delta t)$ to be trace-preserving, so that $\Tr\left(\mathcal{M}_{W_k} \, \rho \right) = \Tr\left(\rho\right)$, for all $\rho$. However, it is also common to use trace-non-increasing maps, such that $\Tr \left(\mathcal{M}_{W_k} \, \rho\right) = p(W_k|\rho)$~\cite{PRXQuantum.5.020201}.}.	
	
Crucially, the decomposition in Eq.~\eqref{eq:cont_mon_divisible_evolution} is only possible because of the CP-divisibility of the dynamics of a suitably augmented S--RC unit. In this sense, the RC modes absorb all the relevant memory effects, enabling an efficient description of the conditional dynamics of the system. As we discuss below, it is precisely this CP-divisibility property which allows us to estimate unknown parameters within the Bayesian formalism~\cite{PhysRevA.87.032115, PhysRevA.94.032103}.
\newline

\noindent{\it Gaussian continuous measurements---}To fix ideas, let us address the simple problem of estimating an unknown parameter by continuously monitoring a linear bosonic system strongly (and linearly) coupled to a bosonic bath. 
This can be described by the  paradigmatic quantum Brownian motion model~\cite{weiss2012quantum},
\begin{equation}\label{eq:total_ham_original}
    H = H_{\rm S} + H_{\rm B} + H_{\rm SB}.
\end{equation}
Here, $H_{\rm S} = \frac{1}{2} \left[ p_{\rm S}^2 + (\Omega_{\rm S}^2 + \delta \Omega_{\rm S}^2) x_{\rm S}^2 \right]$ describes a harmonic oscillator---the Brownian particle---with frequency $\Omega_{\rm S}$, which is coupled, arbitrarily strongly, to a bosonic bath at temperature $T$ with Hamiltonian $H_{\rm B} = \textstyle{\frac{1}{2} \sum_j (p_j^2 + \omega_j^2 x_j^2)}$. The interaction between the system and the bath is modeled by a position--position coupling term, $H_{\rm SB} = -x_{\rm S} \textstyle{\sum_j c_j x_j}$. Throughout the text, we assume unit masses for both system and bath modes. We further define the bath spectral density as
\begin{equation}\label{SDdiracdelta}
    \mathcal{J} (\omega) \coloneqq \dfrac{\pi}{2} \sum_j \dfrac{c_j^2}{\omega_j} \delta (\omega - \omega_j).
\end{equation}
The interaction Hamiltonian introduces a frequency-renormalization shift to the system, $\delta \Omega_{\rm S}^2$, which can be written in terms of the spectral density as follows~\cite{caldeira1983quantum,weiss2012quantum}:
\begin{equation}
    \delta \Omega_{\rm S}^2 \coloneqq \sum_j \dfrac{c_j^2}{\omega_j^2} = \dfrac{2}{\pi} \int_0^\infty \dd{\omega} \dfrac{\mathcal{J} (\omega)}{\omega}.
\end{equation}

As detailed in the next section, the reaction coordinate mapping preserves linearity. Measurements, on the contrary, can yield non-Gaussian conditional states, in general. We, therefore, restrict ourselves to Gaussian continuous measurements, such as homodyne detection~\cite{serafini2023quantum,RevModPhys.84.621, DeSosa2025PRECoolingWithContinuousFeedback}, which are among the most common in the toolbox of quantum-optical platforms.
Let us take $R = [x_{\rm S}, p_{\rm S}, x^1_{\rm R}, p^1_{\rm R}, \dots, x^n_{\rm R}, p^n_{\rm R}]^T$ to be the vector of quadrature operators of the S--RC unit, with $n$ RC modes (see below for a derivation). Here and in what follows, we set $\hbar = k_B = 1$. The quadratures fulfill the canonical commutation relations $\comm{R_j}{R_k} = i\Omega_{jk}$, with $\Omega = \oplus_{l=1}^{n+1}\left[\begin{smallmatrix}
  0& 1\\
  -1& 0
\end{smallmatrix}\right]$. If the system and the bath are initialized in a Gaussian state, linearity guarantees that the S--RC unit can be fully characterized by its first and second moments, $d \in {\mathbb{R}}^{2n+2}$ and $\sigma \in {\mathbb{R}}^{(2n+2) \times (2n+2)}$, respectively, defined as~\cite{RevModPhys.84.621}
\begin{align}
    d_j & \coloneqq {\rm Tr} [\rho\,R_j],\nonumber\\
    \sigma_{jk} & \coloneqq \frac{1}{2}{\rm Tr} [\rho\,\{R_j, R_k\}] - d_j d_k, 
\end{align}
with $\{\cdot,\cdot\}$ denoting the anticommutator. The covariance matrix (CM) respects the Heisenberg uncertainty relation $\sigma \pm i\,\Omega/2 \geq 0$. 

Without considering the measurement outcome, the Gaussian CP-divisible (Markovian) dynamics of the S--RC unit can be described by a quantum dynamical semigroup, whose infinitesimal time updates are given by (see Appendices \ref{app:Gaussian_meas} and \ref{sec:app_Exact_and_GKLS} or~\cite{RevModPhys.84.621})
\begin{align}\label{eq:Gaussian_map}
    d(\theta) & \mapsto  d(\theta) + \delta t\,A({\theta})\,d(\theta), \nonumber\\
    \sigma(\theta) & \mapsto \sigma(\theta) + \delta t\left(A({\theta}) \sigma(\theta) + \sigma(\theta)\,A^T({\theta}) + D({\theta})\right).
\end{align}
Here, $\theta$ is the unknown parameter to be estimated and $A({\theta}) = A_1({\theta}) + A_2$ is the drift matrix, with $A_1({\theta}), A_2 \in \mathbb{R}^{(2n+2)\times(2n+2)}$ being the contributions from the sample (i.e., the bath that encodes the parameter) and the detector environment (i.e., the bath where the detector is located), respectively. In turn, $D({\theta}) = D_1({\theta}) + D_2$ is the diffusion matrix, with $D_1({\theta}), D_2 \in \mathbb{R}^{(2n+2)\times(2n+2)}$ similarly denoting the contributions from the bath and the detector environment. 

The specific expressions for $A_1({\theta})$ and $D_1({\theta})$ depend on the original system--bath Hamiltonian (see Appendix~\ref{app:GKLSME}). For the detector environment contribution, we set $A_2 = -(\lambda / 2) \mathds{1}_{\rm S} \oplus \mathds{O}_{\rm R}$, where the subindices S and R denote the system and the RC modes, respectively. Throughout the paper, we use the notations $\mathds{1}$ and $\mathds{O}$ for the identity and zero array (not necessarily square). In each case, array dimensions will be clear from the context. Here, $D_2 = \lambda\,\sigma_{\rm env} \oplus \mathds{O}_{\rm R}$, where $\sigma_{\rm env}$ is the covariance matrix of the modes in the detector environment. It is common to take $\sigma_{\rm env} = \mathds{1}/2$ and $d_{\rm env} = \mathds{O}$, which corresponds to the vacuum state. However, the formalism is valid for any general Gaussian state. 

Incorporating the outcomes requires specifying the measurement. The Gaussian measurement can be represented by a CM $\sigma^M \in \mathbb{R}^{2\times 2}$ satisfying the Heisenberg uncertainty relation. Common choices of Gaussian measurements are homodyne detection of the $x_{\rm S}$ quadrature, with associated covariance matrix $\sigma^M_{\rm hom} = \lim_{r\to 0}\frac{1}{2}\left[\begin{smallmatrix}
  r& 0\\
  0& 1/r
\end{smallmatrix}\right]$, and heterodyne detection, for which $\sigma^M_{\rm het} = \mathds{1}/2$.  Conditioned on having observed a sequence of outcomes ${\vv W}_{m-1}$, with $W_k\in \mathbb{R}^2$, the Gaussian state of the S--RC unit can be characterized by (see Appendix~\ref{app:Gaussian_meas})
\begin{align}
    d(\theta, {\vv W}_{m-1}) & =[d_{\rm S}(\theta, {\vv W}_{m-1}); d_{\rm R}(\theta, {\vv W}_{m-1})],\nonumber\\
    \sigma(\theta, {\vv W}_{m-1})  & = \left[\begin{matrix}
\sigma_{\rm S}(\theta, {\vv W}_{m-1}) & C(\theta, {\vv W}_{m-1}) \\ C^T(\theta, {\vv W}_{m-1}) & \sigma_{\rm R}(\theta, {\vv W}_{m-1})
    \end{matrix}
    \right].
\end{align}
At step $m$, the Gaussian measurement yields a normally distributed outcome $W_m$,
\begin{align}\label{eq:normal_dist}
    p(W_m|{\vv W}_{m-1}, \theta) = {\cal N}[\xi(\theta, {\vv W}_{m-1}), V(\theta, {\vv W}_{m-1})],
\end{align}
with mean $\xi(\theta, {\vv W}_{m-1}) = \sqrt{\lambda\,\delta t}\,L\,d (\theta, {\vv W}_{m-1})$ and covariance $V({\theta}, {\vv W}_{m-1}) = \sigma_{\rm env} + \lambda\,\delta t [\sigma_{\rm S}({\theta}, {\vv W}_{m-1}) - \sigma_{\rm env}] + \sigma^M$ (see Appendix~\ref{app:Gaussian_meas} and Refs.~\cite{brask2021gaussian,serafini2023quantum} for details). Here, $L=\left[\begin{smallmatrix}
    1 & 0 & 0 & \dots & 0\\0 & 1 & 0 & \dots & 0
\end{smallmatrix}\right]$ extracts the system displacement from the displacement vector of the S--RC unit; that is, $d_{\rm S}(\theta, {\vv W}_{k}) = L\,d(\theta, \vv{W}_k)$. To leading order in $\delta t$, the covariance of the measurement outcome reduces to $V(\theta, {\vv W}_{m-1}) \approx \sigma_{\rm env} + \sigma^M = V$. Conditioned on the outcome $W_m$, the S--RC unit will then evolve according to the quantum Kalman filter~\cite{PhysRevLett.91.250801,Zhang2017,PhysRevLett.91.250801,wiseman_milburn_2009,jacobs_2014}
\begin{align}\label{eq:Wiener_update_General}
    d(\theta, {\vv W}_{m}) & = d(\theta, {\vv W}_{m-1}) + \delta t\,A(\theta)\, d(\theta, {\vv W}_{m-1}) + \sqrt{\lambda\,\delta t} \nonumber\\
    &\times K(\theta, {\vv W}_{m-1})\,[W_m - \sqrt{\lambda\,\delta t}\,d_{\rm S}(\theta, {\vv W}_{m-1})],\nonumber\\
    \sigma(\theta, {\vv W}_m) & = \sigma(\theta, {\vv W}_{m-1}) + \delta t\,[D(\theta) + A(\theta)\,\sigma(\theta, {\vv W}_{m-1}) \nonumber\\
    & \hspace{-1.6cm}+ \sigma(\theta, {\vv W}_{m-1})\,A^T(\theta) - \lambda K(\theta, {\vv W}_{m-1})\,V\,K^T(\theta, {\vv W}_{m-1})],
\end{align}
where we have defined the Kalman gain matrix, $K(\theta, {\vv W}_{m-1})$, as
\begin{equation}
    K(\theta, {\vv W}_{m-1}) \coloneq \left[
    \begin{matrix}
        \sigma_{\rm S}(\theta, {\vv W}_{m-1}) - \sigma_{\rm env}
        \\
        C^T(\theta, {\vv W}_{m-1})
    \end{matrix}\right]
    V^{-1}.
\end{equation}

Eqs.~\eqref{eq:normal_dist} and \eqref{eq:Wiener_update_General} provide an iterative stochastic method to describe the conditional time evolution of the S--RC unit which, in turn, accounts for the continuous monitoring of the original system under general, possibly strongly non-Markovian, dissipation.
\newline

\noindent{\it Setting up the reaction coordinates---}To illustrate our proposed method, we provide details of the RC mapping for the paradigmatic Brownian motion problem in Eq.~\eqref{eq:total_ham_original}.
The Brownian particle is continuously monitored to infer some property of the bath, namely, its temperature. 
In what follows, we take a highly structured Lorentzian profile for the spectral density,
\begin{equation}\label{SDlorentzian}
    \mathcal{J} (\omega) = \dfrac{\gamma g^2 \omega}{\gamma^2 \omega^2 + (\omega^2 - \Omega_\mathrm{R}^2)^2}.
\end{equation}
This choice describes a narrow-band environment peaked around $\Omega_{\rm R}$, with $\gamma$ controlling the bath memory time and $g$ the system-bath coupling strength \cite{Iles_PRA2014, Strasberg_2016}.

We initialize the system in a Gaussian state uncorrelated from the sample bath and in thermal equilibrium. The free dissipative dynamics of the Brownian particle is then exactly solvable by means of the quantum Langevin equation (see Appendix~\ref{app:exactsolution}). 

Starting from the original Hamiltonian in Eq.~\eqref{eq:total_ham_original}, we now sketch how to extract one reaction coordinate from the bath (see details in Appendix~\ref{app:RCmapping}). A reaction coordinate is defined as a collective bath mode
\begin{equation}\label{XRC}
    X_{\rm R} \coloneqq \dfrac{1}{g} \sum_j c_j x_j,
\end{equation}
with $g^2 = \sum_j c_j^2$. Thus, the Hamiltonian can be rewritten in terms of the RC quadratures:
\begin{align}\label{eq:RC_Hamiltonian}
    H &= H_{\rm S} + \dfrac{1}{2} (P_{\rm R}^2 + \Omega_{\rm R}^2 X_{\rm R}^2) - g x_{\rm S} X_{\rm R} - X_{\rm R} \sum_{\alpha \neq \rm R} C_\alpha X_\alpha \notag \\
    &+ \dfrac{1}{2} \sum_{\alpha \neq \rm R} (P_\alpha^2 + \Omega_\alpha^2 X_\alpha^2),
\end{align}
where $X_{\rm R}$ and $P_{\rm R}$ are the canonical quadratures of the reaction coordinate, while $X_{\alpha}$ and $P_{\alpha}$ are the quadratures of the residual bath. Here, the coefficients $C_\alpha$ and residual mode frequencies $\Omega_\alpha$ are set by the canonical transformation defining the RC mapping. The mapping fixes both the RC frequency, $\Omega_{\rm R}^2 = \frac{1}{g^2} \sum_j c_j^2 \omega_j^2$, and the residual bath spectral density, which for our choice in Eq.~\eqref{SDlorentzian} is $\mathcal{J}_{\rm R} (\omega) = \gamma \omega e^{-\abs{\omega} / \Lambda}$. This is strictly Ohmic in the limit of $\Lambda \to \infty$, with $\Lambda$ being the cut-off frequency. Note also that $\Omega_{\rm R}$ denotes the bare RC frequency fixed by the mapping, whereas in Eq.~\eqref{SDlorentzian} it denotes the corresponding renormalized frequency; see Appendix~\ref{app:RCmapping} for details.

Should the dissipative dynamics of the resulting S--RC unit still not admit a Markovian description, one may iterate the procedure to extract more reaction coordinates. In appropriate parameter ranges, the augmented system thus built is eventually described by a global GKLS master equation\footnote{A global (rather than local) approach is necessary, since the interaction between the system and the RC can be strong, due to a strong coupling between the bare system and the original bath. See, e.g., Refs.~\cite{hofer2017markovian,gonzalez2017testing} and references therein for a discussion on local/global master equations.}. In our example, extracting a single reaction coordinate mode is sufficient (see Appendix~\ref{sec:app_Exact_and_GKLS}). We cast the dissipative dynamics of our simple S--RC unit in the form of Eq.~\eqref{eq:Gaussian_map}, where we analytically compute the associated drift and diffusion matrices and capture their dependence on dissipation rates, temperature and frequencies.
\newline

\noindent{\it Example: Bayesian continuous thermometry---}We have now provided all the key ingredients needed for the estimation of any bath parameter from a continuous measurement record. Taking the limit of discrete time steps, upon observing a sequence of measurement outcomes $\vv{W}_m = \{W_m,\dots,W_1\}$, and starting from a prior knowledge $p(\theta)$, one may use Bayes' update rule to obtain the posterior
\begin{align}\label{eq:bayes}
    p(\theta | {\vv W}_m) 
    & = \frac{p({\vv W}_m|\theta) p(\theta)}{p({\vv W}_m)} = \dfrac{\Pi_{k=1}^m p(W_k|{\vv W}_{k-1}, \theta) p(\theta)}{\Pi_{k=1}^m p(W_k|{\vv W}_{k-1})} \nonumber\\
    & = \dfrac{p(W_m|{\vv W}_{m-1}, \theta) p(\theta|{\vv W}_{m-1})}{p({W}_m|{\vv W}_{m-1})}.
\end{align}

\begin{figure}
    \centering
    \includegraphics[width=.99\linewidth]{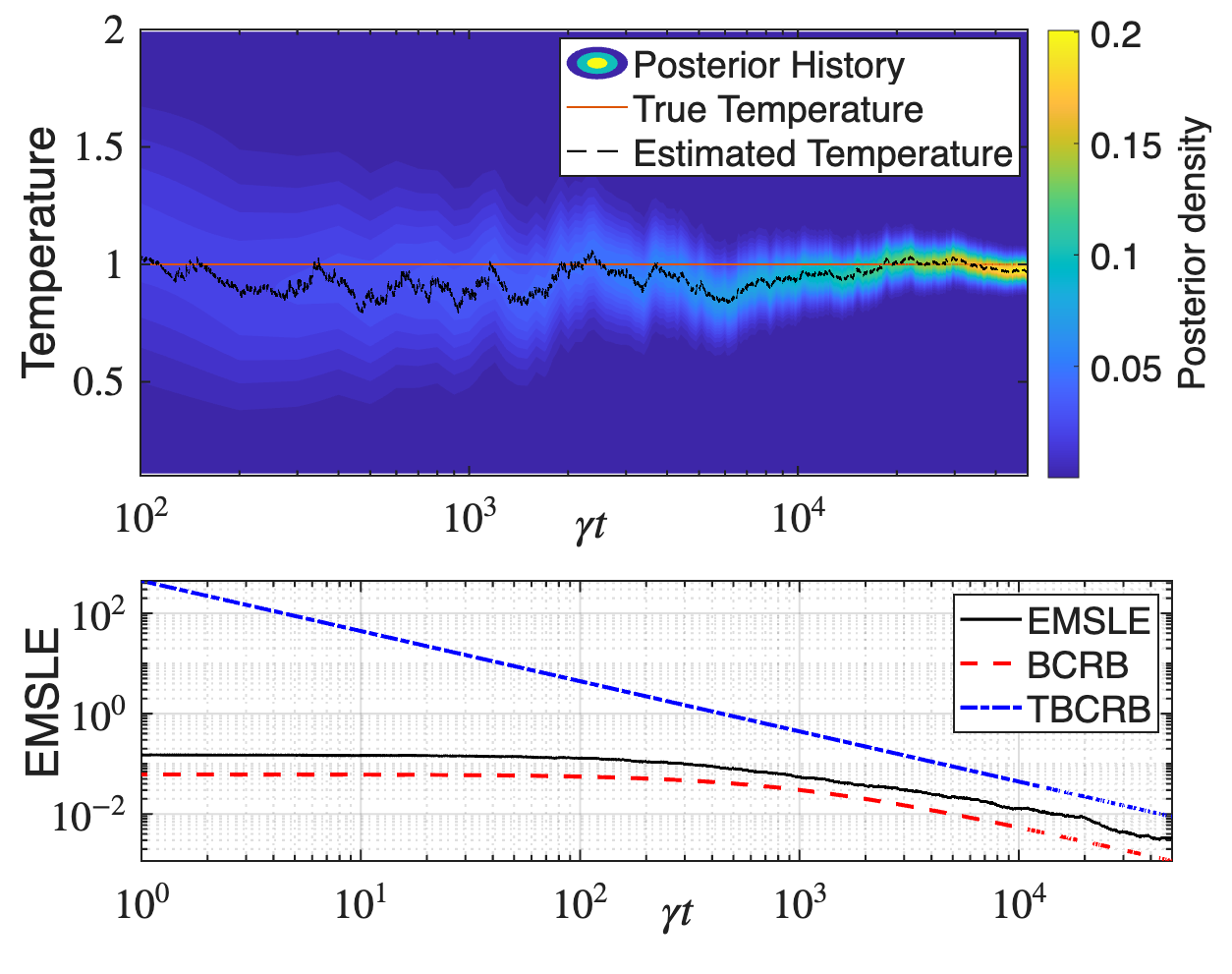}
    \caption{Top: A sample Monte Carlo simulation of the homodyne outcomes is used for Bayesian inference. The contour plot shows the posterior density function conditioned on the observed data. Here, the red and black lines indicate the true and estimated temperatures, respectively. As more data are collected, the posterior becomes sharper and the estimator is more precise (on average). {Bottom:} We estimate the EMSLE using $320$ Monte Carlo simulations. We use the Fisher information to lower bound the EMSLE using the Bayesian Cram\'er-Rao bound (red dashed). We also use the FI to evaluate the tight Bayesian Cram\'er-Rao bound (blue dashed-dotted). The bound holds \textit{only} for unbiased estimators. Thus, it is tight and saturable only asymptotically.  
    The system parameters are $\gamma = \Omega_{\rm S}$, $g = \sqrt{10}$, $\Lambda = 1000\Omega_{\rm S}$, $\lambda = \Omega_{\rm S}$, $\Omega_{\rm R} = \sqrt{10}$, and $\Omega_{\rm S} = 1$. For the prior, we set $T_{\min}=0.1$, $T_{\max}=2$, and $\alpha=0$.}
    \label{fig:Bayesian_estimation}
\end{figure}

Here, $p({\vv W}_m) = \int \dd{\theta} p({\vv W}_m|\theta) p(\theta)$ and $p(W_k|{\vv W}_{k-1}) = \int \dd{\theta} p(W_k|{\vv W}_{k-1}, \theta) p(\theta|{\vv W}_{k-1})$ are normalization factors, with $p({W}_k|{\vv W}_{k-1}, \theta)$ being the likelihood of observing outcome $W_k$ conditioned on the parameter being $\theta$ and the measurement record history ${\vv W}_{k-1}$. In particular, $p(W_1|{\vv W}_{0}, \theta) = p({W}_1|\theta)$. In our example, this likelihood is drawn from Eq.~\eqref{eq:normal_dist}. The last equality in Eq.~\eqref{eq:bayes} allows for an iterative update of the posterior distribution as new data are obtained. Furthermore, Eqs.~\eqref{eq:Wiener_update_General} provide the update rules for the (conditional) input state, which will then be used to compute the likelihood for the next shot. At any given time step, the posterior distribution can be used to assign an estimate to the unknown parameter. 

To proceed further, let us take the specific example of thermometry, i.e., taking the temperature of the sample bath as the unknown parameter $\theta$ (that is, $\theta \equiv T$), and fix homodyne detection of the system's $x_{\rm S}$ quadrature as our Gaussian measurement. We further assume, without loss of generality, that the detector environment is at zero temperature, and take the initial smooth prior
\begin{align}
    p(T) = \frac{e^{\alpha \sin^2\left[\pi\tfrac{T-T_{\min}}{T_{\max}-T_{\min}}\right]} - 1}{(T_{\max}-T_{\min})[e^{\alpha/2}I_0(\alpha/2)-1]},
\end{align}
with $I_0$ being the modified Bessel function of the first kind, and $T_{\min}$ and $T_{\max}$ mark the lower and upper domain of the prior. Here, we take the expected mean square logarithmic error (EMSLE) \cite{PhysRevLett.127.190402,Rubio2022} as the loss function to be minimized in searching for an estimate, i.e.,
\begin{equation}\label{eq:ERMSE}
\hspace{-.2cm}
    {\rm EMSLE} \hspace{-.05cm} = \hspace{-.1cm}\int \hspace{-.1cm}\dd{T} p(T) \hspace{-.1cm}\int \hspace{-.1cm}\dd{\vv W}_{\hspace{-.1cm} m} p({\vv W}_{\hspace{-.1cm}m}|T) \log^2 \hspace{-.1cm} \left[\frac{{\hat T}({\vv W}_{\hspace{-.1cm}m})}{T}\right],
\end{equation}
although our ideas are applicable to any other figure of merit, namely, the expected [relative] mean square error \cite{PhysRevA.104.052214}. For our choice, the optimal estimator $T^*$ is known to be~\cite{PhysRevLett.127.190402}
\begin{equation}\label{eq:bayesian-estimator}
    T^*({\vv W}_{\hspace{-.1cm}m}) = \exp\left[\int \dd{T} p(T|\vv W_{\hspace{-.1cm}m}) \log (T)\right].
\end{equation}

The exact evaluation of the EMSLE is unmanageable, since the number of potential trajectories grows exponentially with the total measurement time. Instead, one can simulate the trajectories via Monte Carlo techniques, which leads to stable estimates of the EMSLE. To this end, a `true' temperature is randomly sampled from the prior distribution $p(T)$, which will be used to generate the data ${\vv W}_m$ according to the likelihood $p({\vv W}_m|T)$. The temperature estimator is constructed from the simulated data, and the corresponding error is calculated. By repeating this procedure many times and averaging the resulting errors, an estimate of the EMSLE is obtained.

In Fig.~\ref{fig:Bayesian_estimation} we illustrate such an estimation task. Namely, we consider a Brownian probe undergoing strongly non-Markovian dissipation, due to the highly structured spectral density in Eq.~\eqref{SDlorentzian}, and apply the RC mapping to it. We are then able to exploit the (inherently Markovian) procedure from Eqs.~\eqref{eq:Wiener_update_General} in the resulting augmented S--RC unit to construct the necessary posterior distribution in Eq.~\eqref{eq:bayes}, and accurately estimate the temperature of the sample within the Bayesian formalism. The top panel shows the posterior distribution over time, overlaid with the true and estimated temperatures. The bottom panel depicts the EMSLE on a logarithmic time scale, which drops with time as more data are collected. A comparison with the asymptotic bounds found by means of the Fisher information is also performed. While the Bayesian Cram\'er-Rao bound (BCRB) sets a strict lower bound on the error of any estimator, the tight Bayesian Cram\'er-Rao bound (TBCRB) sets a lower bound on the error of \textit{unbiased} estimators but it can be saturated asymptotically. See the End Matter for the derivation of the FI and the CRBs.
\newline

\noindent{\it Conclusion---}We have addressed the open problem of parameter estimation in continuously monitored systems under CP-indivisible non-Markovian dynamics. CP-indivisibility is a major obstacle that precludes the usual sequential update of the conditional state of the probe for interpreting and post-processing a continuous measurement record. We have surmounted this difficulty by resorting to the reaction coordinate mapping; namely, we have proposed the sequential extraction of a number of collective coordinates from the sample bath, replacing the bare probing system with an augmented unit which encapsulates all dissipative memory effects. Once such a unit is found to be weakly coupled to the remaining residual unstructured sample bath, the augmented probe exhibits CP-divisible dynamics which, in turn, allows for the application of standard Markovian data analysis techniques. This allows us to interpret the noisy signal from the continuous monitoring of any probe---regardless of non-Markovianity---to infer unknown parameters of the sample. 

To illustrate the usefulness of our construction, we have estimated the temperature of an equilibrium sample by performing continuous homodyne measurements on a Brownian probe undergoing strongly non-Markovian dissipation. By combining the Bayesian parameter estimation framework with Monte Carlo simulations, we have shown how our approach allows for the conditional updating of the probe state and, eventually, for the construction of accurate temperature estimates. Markovian embedding techniques can thus be a powerful tool enabling quantum parameter estimation from continuous noisy signals whenever the probe--sample interactions are strong and/or highly structured, regardless of the parameter being estimated or the experimental platform of interest. To complement our work, in the End Matter we discuss the asymptotic behavior where we analytically obtain the Fisher information for the continuously monitored system under any Gaussian measurement. 

A natural direction for future work would be to optimize the estimation protocol itself, including the choice of Gaussian measurement, the state (and temperature) of the environmental mode being monitored, and the choice of bath parameters such as the decay rate $\gamma$. 
\newline

\noindent{\it Acknowledgments---}We are thankful to Philipp Stammer and Alessandra Colla for fruitful discussions about the reaction coordinate mapping. The authors acknowledge TU Wien Bibliothek for financial support through its Open Access Funding Programme.
This research was funded in part by the Austrian Science Fund (FWF) [“NOQUS,” Grant DOI: 10.55776/PAT1969224]. P.B. also acknowledges that financial support was provided by the Austrian Science Fund (FWF) through the Stand-Alone grant P 35810-N and P 36633-N. L.A.C. acknowledges support from Ministerio de Ciencia e Innovación and
European Union (FEDER) (PID2022-138269NB-I00) and Ramón y Cajal Fellowship
(RYC2021325804-I), funded by MCIN/AEI/10.13039/501100011033 and ‘NextGenerationEU’/PRTR. P.P.P. acknowledges support from the Swiss National Science Foundation (Eccellenza Professorial Fellowship PCEFP2\_194268).\newline

\noindent{\it Data availability---}No data was collected for this work. The codes for the Monte Carlo simulations resulting in Fig.~\ref{fig:Bayesian_estimation} are publicly available in \href{https://github.com/Mehboudi/NonMarkovian_continuous_monitoring}{this repository}.

\bibliography{Refs}
\pagebreak

\appendix
\onecolumngrid
\section*{End Matter}
\subsection*{The Fisher Information of Gaussian Continuous Measurements and the Asymptotic Behavior}\label{app:FI}
In this End Matter, we provide the explicit derivation of the ultimate precision bounds for parameter estimation discussed in the main text. By tracking how classical information accumulates within the Gaussian continuous measurement record via the filtered stochastic updates, we evaluate the total Fisher information. Our main focus is on the asymptotic limit of large data, that is, $t\to \infty$. For the short-time behavior, see previous results in Refs.~\cite{PhysRevA.95.012116,yokomizo2026fundamentallimitscontinuousgaussian}.

As established in the main text, we consider a continuously monitored Gaussian system characterized by its conditional first and second moments $\{d(\theta, {\vv W}_{m-1}), \sigma(\theta, {\vv W}_{m-1})\}$. Conditioned on the measurement outcome history ${\vv W}_{m-1} = \{W_{m-1},\dots, W_1\}$, the probability distribution for outcome $W_m$ at step $m$ takes the Gaussian form $p(W_m|{\vv W_{m-1}}, \theta) = {\cal N}(\xi(\theta, {\vv W}_{m-1}),V(\theta, {\vv W}_{m-1}))$, with conditional mean $\xi(\theta, {\vv W}_{m-1}) = \sqrt{\lambda \delta t} L d(\theta, {\vv W}_{m-1})$ and variance $V(\theta, {\vv W}_{m-1}) = \sigma_{\rm env} + \lambda\,\delta t \left[ \sigma_{\rm S}(\theta, {\vv W}_{m-1}) - \sigma_{\rm env} \right] +\sigma^M$. Here, $L=\left[\begin{smallmatrix} 1 & 0 & 0 & \dots & 0\\0 & 1 & 0 & \dots & 0 \end{smallmatrix}\right]$ is the projection matrix mapping the S--RC unit moments to the observed system displacement via $d_{\rm S}(\theta, \vv W_{m-1}) = L d(\theta, \vv W_{m-1})$.

Let us restrict ourselves to the case where we ignore the term $\lambda \delta t$ in the variance, which is valid for $\lambda \delta t\ll 1$, and hence replace $V(\theta, {\vv W}_{m-1}) \approx \sigma_{\rm env}+\sigma^M = V$. Furthermore, we consider a scenario in which the system has reached a stationary state, that is, its covariance matrix no longer evolves. Conditioned on outcome $W_m$, the state is updated to  
\begin{align}
    d(\theta, {\vv W}_{m-1}) & \mapsto (\mathds{1} + \delta t A(\theta))d(\theta, {\vv W}_{m-1}) + \sqrt{\lambda}K(\theta)  Y(\theta, \vv W_m)  = d(\theta, {\vv W}_m) \text{~~with~~}K(\theta) = \left[
    \begin{matrix}
        \sigma_{\rm S}(\theta) - \sigma_{\rm env}
        \\
        C^T(\theta)
    \end{matrix}\right]
    V^{-1}\nonumber\\
    \sigma(\theta) & \mapsto \sigma(\theta),~~\text{which is the solution to~~}A(\theta) \sigma(\theta) + \sigma(\theta) A^T(\theta) + D(\theta) - \lambda K(\theta) V K^T(\theta) = 0.
\end{align}
Here, we have defined $Y(\theta, \vv W_m) \coloneqq \sqrt{\delta t} W_m - \delta t \sqrt{\lambda} d_{\rm S}(\theta, \vv W_{m-1})$, which is a Wiener process with a vanishing mean and a variance that is proportional to $\delta t V$. Note that the observed data are ${\vv W_m}$ and not $Y(\theta, {\vv W}_m)$. 

To obtain the Fisher information of the entire data, we note that the data themselves are not necessarily independent. For instance, we have $p({\vv W}_{n}|\theta) = p(W_{n}|{\vv W}_{n-1}, \theta) p({\vv W}_{n-1}|\theta)= \prod_{k=1}^n p(W_{k}|{\vv W}_{k-1}, \theta)$, where ${\vv W}_0$ corresponds to no data, i.e., $p(W_1|{\vv W}_0, \theta) \equiv p(W_1|\theta)$. The FI can then be cast as
\begin{align}
    F_{\rm total}(\theta) & = \left\langle [\partial_{\theta} \log p({\vv W}_n|\theta)]^2\right\rangle
    = \int \dd{\vv W}_n p({\vv W}_n|\theta) \left[\sum_{m=1}^n \partial_{\theta} \log p(W_m|{\vv W}_{m-1}, \theta)\right]^2\nonumber\\
    & = \sum_{m=1}^n\int \dd{\vv W}_n p({\vv W}_n|\theta) \left[\partial_{\theta} \log p(W_m|{\vv W}_{m-1}, \theta)\right]^2 \text{~~since the cross terms disappear}\nonumber\\
    & = \sum_{m=1}^n\int \dd{\vv W}_m p({\vv W}_m|\theta) \left[\partial_{\theta} \log p(W_m|{\vv W}_{m-1}, \theta)\right]^2 \text{~~since we can integrate the events after $W_m$ to get one}\nonumber\\
    & = \sum_{m=1}^n\int \dd{\vv W}_{m-1} p({\vv W}_{m-1}|\theta) \int \dd{ W}_m p(W_m|{\vv W}_{m-1}, \theta) \left[\partial_{\theta} \log p(W_m|{\vv W}_{m-1}, \theta)\right]^2\nonumber\\
    & = \sum_{m=1}^n \int \dd{\vv W}_{m-1} p({\vv W}_{m-1}|\theta) F_m({\vv W}_{m-1},\theta)\nonumber\\
    & = \sum_{m=1}^n F_m(\theta),~~\text{with the average FI gain~~} F_m(\theta) \coloneqq \int \dd{\vv W}_{m-1} p({\vv W}_{m-1}|\theta) F_m({\vv W}_{m-1},\theta),\nonumber\\
    & \hspace{1.3cm}\text{and with the conditional FI~~} F_m({\vv W}_{m-1},\theta) \coloneqq \int \dd{ W}_mp(W_m|{\vv W}_{m-1}, \theta) \left[\partial_{\theta} \log p(W_m|{\vv W}_{m-1}, \theta)\right]^2.
\end{align}
Since we are dealing with Gaussian distributions (i.e., the variance is independent of the parameter to leading order, but the mean depends on it), the average FI gain from the $m$-th observation can be alternatively written as~\cite{Malag2015,Ober2002,navarro2025super}
\begin{align}
    F_m(\theta) & = (\lambda \delta t){\mathbb{E}}\left[\partial_{\theta}d(\theta, {\vv W_m})^T L^T V^{-1} L \partial_{\theta} d(\theta, {\vv W_m})\right]_{\vv W_m}  = (\lambda \delta t) {\rm Tr}\left[{\mathbb{E}}\left[\partial_{\theta}d(\theta, {\vv W_m})^TL^T V^{-1} L \partial_{\theta}d(\theta, {\vv W_m})\right]_{\vv W_m}\right] \nonumber\\
    & 
    = (\lambda \delta t) {\mathbb{E}}\left[{\rm Tr}\left[\partial_{\theta}d(\theta, {\vv W_m})^T L^TV^{-1} L \, \partial_{\theta}d(\theta, {\vv W_m})\right]\right]_{\vv W_m} 
    = (\lambda \delta t) {\mathbb{E}}\left[{\rm Tr}\left[L^TV^{-1} L \, \partial_{\theta}d(\theta, {\vv W_m})\partial_{\theta}d(\theta, {\vv W_m})^T\right]\right]_{\vv W_m}
    \nonumber\\
    & = (\lambda \delta t) {\rm Tr}\left[L^TV^{-1} L \, {\mathbb{E}}\left[\partial_{\theta}d(\theta, {\vv W_m})\partial_{\theta}d(\theta, {\vv W_m})^T\right]_{\vv W_m}\right].
\end{align} 
Summing over $m$, we obtain the total FI:
\begin{align}
    F_{\rm total}(\theta) & = \sum_{m=1}^n F_m(\theta) =
    \lambda \delta t \sum_{m=1}^{n}{\rm Tr}\left[L^TV^{-1} L{\mathbb{E}}\left[\partial_{\theta}d(\theta, {\vv W_m})\partial_{\theta}d(\theta, {\vv W_m})^T\right]_{\vv W_m}\right].
\end{align}
Since $\delta t \ll 1$, we can change the summation with an integral. By defining $\tau\coloneqq n\delta t$ and $f(t=m\delta t) \coloneqq  {\mathbb{E}}\left[\partial_{\theta}d(\theta, {\vv W_m})\partial_{\theta}d(\theta, {\vv W_m})^T\right]_{\vv W_m}$ we have
\begin{align}
    F_{\rm total}(\theta) & = \lambda \int_0^{\tau} \dd{t} {\rm Tr}[L^T V^{-1} L f(t)] = \lambda {\rm Tr}\left[L^T V^{-1} L \int_0^{\tau} dt f(t)\right].
\end{align}
We need to evaluate the integral in the large-data limit, i.e., $\tau \gg 1$. To proceed further, note that we have
\begin{align}
    \delta d(\theta, {\vv W_m}) & = d(\theta, {\vv W_m}) - d(\theta, {\vv W}_{m-1})  \nonumber\\
    & = \delta t A(\theta) d(\theta, {\vv W}_{m-1}) + \sqrt{\lambda} K(\theta) Y(\theta, {\vv W}_{m}),\nonumber\\
    \delta \partial_{\theta} d(\theta, {\vv W}_m) & = \partial_{\theta} d(\theta, {\vv W}_{m}) - \partial_{\theta} d(\theta, {\vv W}_{m-1}) \nonumber\\
    &=  \delta t [A(\theta) - \lambda K (\theta) L]\partial_{\theta}d(\theta, {\vv W}_{m-1}) + \delta t A^{\prime} (\theta) d(\theta, {\vv W}_{m-1}) 
    + \sqrt{\lambda}K^{\prime} (\theta) Y(\theta, {\vv W}_{m}),
\end{align}
with primes denoting derivatives with respect to $\theta$. We are interested in calculating $f(t)$ to later evaluate the FI. This function can be found by using the augmented system $Z(\theta, {\vv W}_{m}) = \left[\begin{smallmatrix}
        d(\theta, {\vv W}_{m}) \\
        \partial_{\theta}d(\theta, {\vv W}_{m})
    \end{smallmatrix}\right]$
such that
\begin{align}
    \delta Z (\theta, {\vv W}_{m}) = \delta t {\cal A} Z(\theta, {\vv W}_{m-1}) + {\cal K} Y(\theta, {\vv W}_m), \text{~~with~~} {\cal A} = \left[\begin{matrix}
        A (\theta) & 0\\
        A^{\prime} (\theta) & A (\theta) - \lambda K (\theta) L
    \end{matrix}\right],
    \text{~~and~~} {\cal K} = \left[\begin{matrix}
        \sqrt{\lambda} K (\theta) \\
        \sqrt{\lambda} K^{\prime} (\theta) 
    \end{matrix}\right].
\end{align}
Then, by defining $\Sigma(\theta) = {\mathbb{E}}[Z (\theta, {\vv W}_{m}) Z^T (\theta, {\vv W}_{m})]_{\vv W_m}$ to leading order in $\delta t$, we have 
\begin{align}
    \delta \Sigma(\theta) = \delta t \left({\cal A} \Sigma(\theta) + \Sigma(\theta) {\cal A}^T + {\cal K} V {\cal K}^T\right), 
\end{align}
where we used that $Y(\theta, {\vv W}_{m})$ are i.i.d. with vanishing mean and variance equal to $\delta t V$.
The steady state is the solution to the Lyapunov equation
\begin{align}
    {\cal A} \Sigma^{\rm ss}(\theta) + \Sigma^{\rm ss}(\theta) {\cal A}^T + {\cal K} V {\cal K}^T = 0.
\end{align}
Then, at the steady state, we have $f(t\gg 1) = {\cal Q} {\Sigma^{\rm ss}(\theta)} {\cal Q}^T$, with ${\cal Q} = [\mathds{O},~~\mathds{1}]$ helping to extract the bottom right block of $\Sigma^{\rm ss}(\theta)$. For large signals, we can approximate the FI as follows:
\begin{align}
    F_{\rm total}(\theta) = \tau \lambda {\rm Tr}[L^T V^{-1} L {\cal Q} {\Sigma^{\rm ss}(\theta)} {\cal Q}^T].
\end{align}
\subsection*{Bayesian Cramér-Rao bound (BCRB)}
The FI that we obtained can be used to obtain a BCRB. The relative error can be cast as
\begin{align}
    {\rm EMSLE} = \int \dd{T} p(T) \int \dd{ {\vec W}_n} p({\vec W}_n|T) \log^2 \hspace{-.1cm} \left[\frac{{\hat T}({\vv W}_{\hspace{-.1cm}n})}{T}\right].
\end{align}
This is simply the EMSE for the parameter $\phi = \log T$. Thus, one can use the standard 
Van Trees inequality for $\phi$, and then change the variables back to $T$ to get ~\cite{PhysRevLett.127.190402,PhysRevA.105.042601,PhysRevLett.128.130502}
\begin{align}\label{eq:Van_Trees}
    {\rm EMSLE}^{-1} \leq Q(T) + \int \dd{T} p(T)T^2 F_{\rm total}(T)
\end{align}
with $Q (T)$ being the prior information, given by
\begin{align}
    Q(T) = \int \dd{T} p(T) [T\partial_{T} \log(p(T)) + 1]^2.
\end{align}
While the bound Eq.~\eqref{eq:Van_Trees} holds for any estimator, it may generally be loose. Alternatively, for unbiased estimators, one could use the frequentist Cramér-Rao bound to write~\cite{PhysRevLett.127.190402,PhysRevA.105.042601,PhysRevLett.128.130502}
\begin{align}\label{eq:EMSLE_unbiased}
    {\rm EMSLE}_{\rm unbiased} \gtrsim \int d T \frac{p(T)}{T^2 F_{\rm total}(T)}.
\end{align}
Note that, in the limit of small data, this inequality may not even hold, since
we cannot necessarily find an unbiased estimator. However, at the limit of large data, eventually one can find an unbiased estimator. In that case, the inequality holds, and it is strictly tighter than Eq.~\eqref{eq:Van_Trees}. To confirm, note that at the limit of large data one can ignore the prior information. Then, Eq.~\eqref{eq:Van_Trees} can be obtained from Eq.~\eqref{eq:EMSLE_unbiased} by applying Jensen's inequality. 

\section{Formalism of the Reaction Coordinate Mapping}\label{app:RCmapping}
To ensure our work is self-contained, in this appendix we discuss the details of the reaction coordinate mapping used in the main text, specifically how the Hamiltonian and the spectral density are transformed. We closely follow the derivations in Refs.~\cite{Iles_PRA2014,Strasberg_2016,Nazir2018ReactionCoordinate}.

\subsection{Derivation of the Reaction Coordinate Hamiltonian}

We begin by explicitly writing the original system-bath Hamiltonian
\begin{equation}\label{eq:original_ham_appendix}
    H = \frac{1}{2} \left[ p_{\rm S}^2 + (\Omega_{\rm S}^2 + \delta \Omega_{\rm S}^2) x_{\rm S}^2 \right] + \frac{1}{2} \sum_j (p_j^2 + \omega_j^2 x_j^2) - x_{\rm S} \sum_j c_j x_j.
\end{equation}
Here, the system is a harmonic oscillator with bare frequency $\Omega_{\rm S}$, which interacts with a bosonic bath (i.e., also formed by harmonic oscillators) through a position-position coupling. The term $\delta \Omega_{\rm S}^2$ accounts for the bath-induced frequency-renormalization shift.

To compute the Hamiltonian after the RC mapping, we first note that the reaction coordinate is defined as a collective mode of the bath. Accordingly, we map the $N$ modes of the original bath into a RC plus $N-1$ residual bath modes. As mentioned in the main text, the mapping can be applied iteratively, such that more RC modes are extracted. 

We implement the RC mapping through an orthogonal transformation of the original bath canonical position and momentum operators, $x_j$ and $p_j$---which satisfy the standard commutation relations $\comm{x_j}{p_l} = i \delta_{jl}$---into a new set of collective operators, $X_\alpha$ and $P_\alpha$, according to
\begin{align}
    X_\alpha &= \sum_j O_{\alpha j} x_j, \\
    P_\alpha &= \sum_j O_{\alpha j} p_j.
\end{align}
The transfer matrix $O$ must be identical for both operators to preserve the canonical commutation relations:
\begin{equation}
    \comm{X_\alpha}{P_\beta} = \sum_{j,l} O_{\alpha j} O_{\beta l} \underbrace{\comm{x_j}{p_l}}_{i \delta_{jl}} = i \sum_{j,l} O_{\alpha j} O_{\beta l} \delta_{jl} = i \sum_j O_{\alpha j} O_{\beta j} = i \delta_{\alpha \beta},
\end{equation}
since $O$ is an orthogonal matrix (i.e., $O^\mathrm{T} = O^{-1}$). We have the freedom to choose the transformation such that the residual bath modes also assume a normal form, which implies
\begin{equation}\label{relationfrequencies}
    \sum_j O_{\alpha j} \omega_j^2 O_{\beta j} = \delta_{\alpha \beta} \Omega_\alpha^2,
\end{equation}
with $\alpha,\beta \neq \rm R$, since the row defining the reaction coordinate is fixed by the definition of $X_{\rm R}$. Using these insights, we can rewrite the Hamiltonian as a function of the new operators as
\begin{equation}
    H = H_{\rm S} + \dfrac{1}{2} (P_\mathrm{R}^2 + \Omega_\mathrm{R}^2 X_\mathrm{R}^2) - g x_{\rm S} X_\mathrm{R} - X_\mathrm{R} \sum_{\alpha \neq \rm R} C_\alpha X_\alpha + \dfrac{1}{2} \sum_{\alpha \neq R} (P_\alpha^2 + \Omega_\alpha^2 X_\alpha^2),
\end{equation}
with $C_\alpha = -\sum_j \omega_j^2 O_{{\rm R}j} O_{\alpha j}$. Observing that $O_{{\rm R}j} = c_j / g$, we derive a constraint for $g$ from the canonical commutation relation for the RC operators:
\begin{align}
    \comm{X_\mathrm{R}}{P_\mathrm{R}} &= \sum_{j,l} \comm{O_{{\rm R}j} x_j}{O_{{\rm R}l} p_l} = \sum_{j,l} O_{{\rm R}j} O_{{\rm R}l} \comm{x_j}{p_l} = i \sum_{j,l} O_{{\rm R}j} O_{{\rm R}l} \delta_{jl} = i \sum_j O_{{\rm R}j}^2 = \dfrac{i}{g^2} \sum_j c_j^2 = i,
\end{align}
which yields the following condition:
\begin{equation}
    g^2  = \sum_j c_j^2.
\end{equation}
To determine all transformation parameters exclusively from the original spectral density, we note from Eq.~\eqref{relationfrequencies} that the RC frequency fulfills
\begin{equation}
    \Omega_\mathrm{R}^2 = \sum_j \omega_j^2 O_{{\rm R}j}^2 = \dfrac{1}{g^2} \sum_j c_j^2 \omega_j^2.
\end{equation}
By substituting the spectral density from Eq.~\eqref{SDdiracdelta}, the following relations hold:
\begin{align}
    \delta \Omega_\mathrm{S}^2 &:= \sum_j \dfrac{c_j^2}{\omega_j^2} = \dfrac{2}{\pi} \int_0^\infty \dd{\omega} \dfrac{\mathcal{J} (\omega)}{\omega}, \\
    g^2 &= \dfrac{2}{\pi} \int_0^\infty \dd{\omega} \omega \mathcal{J} (\omega), \label{lambdasquared} \\
    \Omega_\mathrm{R}^2 &= \dfrac{2}{\pi g^2} \int_0^\infty \dd{\omega} \omega^3 \mathcal{J} (\omega). \label{RCfrequency}
\end{align}
It remains to relate $\mathcal{J} (\omega)$ to the spectral density of the RC, $\mathcal{J}_\mathrm{R} (\omega)$. We provide a proof analogous to those in the literature---see also Refs.~\cite{MartinazzoVacchiniHughesBurghardt2011, PhysRevB.30.1208}---based on the requirement that the Fourier-space propagator in both coordinate systems must be identical. The resulting expression, detailed in Appendix \ref{sectionspectraldensities}, is given by
\begin{equation}\label{relationspectraldensities}
    \mathcal{J}_\mathrm{R} (\omega) = \dfrac{g^2 \mathcal{J} (\omega)}{\abs{W^{+} (\omega)}^2},
\end{equation}
where $W^+ (\omega)$ is the boundary value of the Cauchy transform of the original spectral density, defined as
\begin{equation}
    W (z) = \dfrac{1}{\pi} \int_{-\infty}^{+\infty} \dd{\omega} \dfrac{\mathcal{J} (\omega)}{\omega - z},
\end{equation}
with
\begin{equation}
    W^+ (\omega) := \lim_{\epsilon \searrow 0} W(\omega + i \epsilon), \quad \omega \in \mathbb{R},
\end{equation}
where the limit emphasizes that $\omega$ is approached from the upper half of the complex plane. Consequently, the RC spectral density takes the following form:
\begin{equation}\label{frequencyrelation}
    \mathcal{J}_\mathrm{R} (\omega) = \dfrac{\pi}{2} \sum_{\alpha \neq \rm R} \dfrac{C_\alpha^2}{\Omega_\alpha} \delta (\omega - \Omega_\alpha),
\end{equation}
and the following relation also holds:
\begin{equation}\label{physicalRCfrequency}
    \dfrac{g^2}{\delta \Omega_\mathrm{S}^2} = \Omega_\mathrm{R}^2 - \delta \Omega_\mathrm{R}^2,
\end{equation}
with
\begin{equation}\label{deltaOmegaRC}
    \delta \Omega_\mathrm{R}^2 := \sum_{\alpha \neq \rm R} \dfrac{C_\alpha^2}{\Omega_\alpha^2}.
\end{equation}
Note that Eq.~\eqref{physicalRCfrequency} defines the squared physical RC frequency. Using Eq.~\eqref{frequencyrelation}, we can rewrite the total Hamiltonian as 
\begin{equation}\label{finalRChamiltonian}
    H = \dfrac{1}{2} (p_{\rm S}^2 + \Omega_{\rm S}^2 x_{\rm S}^2) + \dfrac{1}{2} \left[ P_\mathrm{R}^2 + \dfrac{g^2}{\delta \Omega_\mathrm{S}^2} \left( X_\mathrm{R} - \dfrac{\delta \Omega_\mathrm{S}^2}{g} x_{\rm S} \right)^2 \right] + \dfrac{1}{2} \sum_{\alpha \neq \rm R} \left[ P_\alpha^2 + \Omega_\alpha^2 \left( X_\alpha - \dfrac{C_\alpha}{\Omega_\alpha^2} X_\mathrm{R} \right)^2 \right]
\end{equation}
which is in Brownian form and reproduces the RC Hamiltonian in Eq.~\eqref{eq:RC_Hamiltonian}.

\subsection{Effective Spectral Density of the Reaction Coordinate}\label{sectionspectraldensities}

In this appendix, we derive the exact relation between the original bath spectral density, $\mathcal{J}(\omega)$, and the effective reaction coordinate spectral density, $\mathcal{J}_{\rm R}(\omega)$. To achieve this, we compute the Fourier-space propagators for the system oscillator in both the original and transformed coordinate frames and require that they yield identical dynamics. 

Starting from the original Hamiltonian in Eq.~\eqref{eq:original_ham_appendix}, the equations of motion for the system and bath position operators evaluate to
\begin{align}
    \Ddot{x}_\mathrm{S} (t) &= -\Omega_\mathrm{S}^2 x_\mathrm{S} (t) - x_\mathrm{S} (t) \sum_j \dfrac{c_j^2}{\omega_j^2} + \sum_j x_j (t) c_j, \label{eq:A20} \\
    \Ddot{x}_j (t) &= -\omega_j^2 x_j (t) + c_j x_\mathrm{S} (t). \label{eq:A21}
\end{align}
Fourier-transforming these equations according to the convention $\Tilde{f} (z) \coloneqq \int_{-\infty}^{+\infty} \dd{t} f(t) e^{i z t}$ yields:
\begin{align}
    -z^2 \Tilde{x}_\mathrm{S} (z) &= -\Omega_\mathrm{S}^2 \Tilde{x}_\mathrm{S} (z) - \Tilde{x}_\mathrm{S} (z) \sum_j \dfrac{c_j^2}{\omega_j^2} + \sum_j \Tilde{x}_j (z) c_j, \\
    -z^2 \Tilde{x}_j (z) &= -\omega_j^2 \Tilde{x}_j (z) + c_j \Tilde{x}_\mathrm{S} (z).
\end{align}
Substituting the second equation into the first one gives:
\begin{equation}
    -\Omega_\mathrm{S}^2 \Tilde{x}_\mathrm{S} (z) = \Tilde{\mathcal{L}}_\mathrm{S} (z) \Tilde{x}_\mathrm{S} (z), \quad \Tilde{\mathcal{L}}_\mathrm{S} (z) \coloneqq -z^2 + \sum_j \dfrac{c_j^2}{\omega_j^2} - \sum_j \dfrac{c_j^2}{\omega_j^2 - z^2}.
\end{equation}
We apply an analogous procedure to the RC Hamiltonian in Eq.~\eqref{eq:RC_Hamiltonian}, which generates the following three equations of motion:
\begin{align}
    \Ddot{x}_\mathrm{S} (t) &= -\Omega_\mathrm{S}^2 x_\mathrm{S} (t) - \delta \Omega_\mathrm{S}^2 x_\mathrm{S} (t) + g X_\mathrm{R} (t), \\
    \Ddot{X}_\mathrm{R} (t) &= -\Omega_\mathrm{R}^2 X_\mathrm{R} (t) + g x_\mathrm{S} (t) + \sum_{\alpha \neq \rm R} C_\alpha X_\alpha (t), \\
    \Ddot{X}_\alpha (t) &= -\Omega_\alpha^2 X_\alpha (t) + C_\alpha X_\mathrm{R} (t).
\end{align}
Fourier-transforming these equations we get:
\begin{align}
    -z^2 \Tilde{x}_\mathrm{S} (z) &= -\Omega_\mathrm{S}^2 \Tilde{x}_\mathrm{S} (z) - \delta \Omega_\mathrm{S}^2 \Tilde{x}_\mathrm{S} (z) + g \Tilde{X}_\mathrm{R} (z), \\
    -z^2 \Tilde{X}_\mathrm{R} (z) &= -\Omega_\mathrm{R}^2 \Tilde{X}_\mathrm{R} (z) + g \Tilde{x}_\mathrm{S} (z) + \sum_{\alpha \neq \rm R} C_\alpha \Tilde{X}_\alpha (z), \\
    -z^2 \Tilde{X}_\alpha (z) &= -\Omega_\alpha^2 \Tilde{X}_\alpha (z) + C_\alpha \Tilde{X}_\mathrm{R} (z),
\end{align}
from which we can extract the system propagator as
\begin{equation}
    -\Omega_\mathrm{S}^2 \Tilde{x}_\mathrm{S} (z) = \Tilde{\mathcal{L}} (z) \Tilde{x}_\mathrm{S} (z), \quad \Tilde{\mathcal{L}} (z) \coloneqq -z^2 + \delta \Omega_\mathrm{S}^2 - \dfrac{g^2}{-z^2 + \Omega_\mathrm{R}^2 - \sum_{\alpha \neq \rm R} \frac{C_\alpha^2}{\Omega_\alpha^2 - z^2}}.
\end{equation}
Since the reaction coordinate mapping corresponds only to a redefinition of environmental coordinates, the reduced dynamics of the system must remain invariant under the RC mapping. Hence, the Fourier-space propagators must be identical in both coordinate systems, and the following relation holds:
\begin{equation}
    -z^2 + \sum_j \dfrac{c_j^2}{\omega_j^2} - \sum_j \dfrac{c_j^2}{\omega_j^2 - z^2} = -z^2 + \delta \Omega_\mathrm{S}^2 - \dfrac{g^2}{-z^2 + \Omega_\mathrm{R}^2 - \sum_{\alpha \neq \rm R} \frac{C_\alpha^2}{\Omega_\alpha^2 - z^2}},
\end{equation}
which gives
\begin{equation}\label{WoW1}
    W_0 (z) = \dfrac{g^2}{-z^2 + \Omega_\mathrm{R}^2 - W_1 (z)},
\end{equation}
where we have defined the Cauchy transform of $\mathcal{J} (\omega)$ (and analogously for the RC spectral density) as
\begin{equation}
    W_0 (z) = \dfrac{1}{\pi} \int_{-\infty}^{+\infty} \dd{\omega} \dfrac{\mathcal{J} (\omega)}{\omega - z},
\end{equation}
which can be inverted to yield
\begin{equation}\label{relationW}
    W_1 (z) = \Omega_\mathrm{R}^2 - z^2 - \dfrac{g^2}{W_0 (z)}.
\end{equation}
Note that we have extended the definition of $\mathcal{J} (\omega)$ to negative frequencies via the odd parity condition $\mathcal{J} (-\omega) = - \mathcal{J} (\omega)$. Eq.~\eqref{relationspectraldensities} is then derived by taking the imaginary part of Eq.~\eqref{relationW} and recognizing that $\mathcal{J} (\omega) = \Im W_0^+ (\omega)$.

\subsubsection{The Ohmic Limit of the RC Spectral Density}
We now verify that the Lorentzian profile in Eq.~\eqref{SDlorentzian} is recovered when imposing an Ohmic profile for the RC spectral density, $\mathcal{J}_{\rm R} (\omega) = \gamma \omega e^{-\abs{\omega} / \Lambda}$, in the limit of an infinite cut-off frequency $\Lambda \to \infty$. We again extend the RC spectral density to negative frequencies via the odd parity condition $\mathcal{J}_{\rm R} (-\omega) = -\mathcal{J}_{\rm R} (\omega)$, which justifies the absolute value inside the exponential cut-off.

We begin by evaluating the boundary value $W_1^+ (\omega)$, which subsequently determines $W_0^+ (\omega)$ via Eq.~\eqref{WoW1}, as
\begin{align}
    W_1^+ (\omega) &= \dfrac{1}{\pi} \lim_{\epsilon \searrow 0} \int_{-\infty}^{\infty} \dd{\omega'} \dfrac{\mathcal{J}_{\rm R} (\omega')}{\omega' - (\omega + i \epsilon)} = \dfrac{1}{\pi} \lim_{\epsilon \searrow 0} \int_{-\infty}^{\infty} \dd{x} \dfrac{\mathcal{J}_{\rm R} (\omega + x)}{x - i \epsilon} \\
    &= \dfrac{1}{\pi} \left( i \pi \int_{-\infty}^{\infty} \dd{x} \mathcal{J}_{\rm R} (\omega + x) \delta (x) + \mathcal{P} \int_{-\infty}^\infty \dd{x} \dfrac{\mathcal{J}_{\rm R} (\omega + x)}{x} \right) \\
    &= i \mathcal{J}_{\rm R} (\omega) + \dfrac{1}{\pi} \mathcal{P} \int_{-\infty}^{\infty} \dd{\omega'} \dfrac{\mathcal{J}_{\rm R} (\omega')}{\omega' - \omega} \coloneqq \Delta_1 (\omega) + i \mathcal{J}_{\rm R} (\omega) \label{eqC18},
\end{align}
where we have used the Sokhotski–Plemelj theorem to evaluate the limit. Focusing specifically on the principal value integral for the Ohmic spectral density, we find
\begin{align}
    \Delta_1 (\omega) &= \dfrac{1}{\pi} \mathcal{P} \int_{-\infty}^{\infty} \dd{\omega'} \dfrac{\gamma \omega' e^{-\abs{\omega'}/\Lambda}}{\omega' - \omega} = \dfrac{1}{\pi} \mathcal{P} \int_{-\infty}^{\infty} \dd{\omega'} \gamma e^{-\abs{\omega'}/\Lambda} \left( 1 + \dfrac{\omega}{\omega'-\omega} \right) \nonumber \\
    &= \dfrac{\gamma}{\pi} \left( \mathcal{P} \int_{-\infty}^\infty \dd{\omega'} e^{-\abs{\omega'}/\Lambda} + \mathcal{P} \int_{-\infty}^\infty \dd{\omega'} \dfrac{\omega e^{-\abs{\omega'}/\Lambda}}{\omega'-\omega} \right) = \dfrac{2 \gamma \Lambda}{\pi} + \dfrac{\gamma}{\pi} \mathcal{P} \int_{-\infty}^\infty \dd{\omega'} \dfrac{\omega e^{-\abs{\omega'}/\Lambda}}{\omega'-\omega} \nonumber \\
    &= \dfrac{2 \gamma \Lambda}{\pi} + \dfrac{\gamma \omega}{\pi} \left( \mathcal{P} \int_0^\infty \dd{\omega'} \dfrac{e^{-\omega'/\Lambda}}{\omega'-\omega} - \mathcal{P} \int_0^\infty \dd{\omega'} \dfrac{e^{-\omega'/\Lambda}}{\omega' + \omega} \right) \nonumber \\
    &= \dfrac{2 \gamma \Lambda}{\pi} - \dfrac{\gamma \omega}{\pi} \left[ e^{-\omega/\Lambda} \Ei \left( \dfrac{\omega}{\Lambda} \right) + e^{\omega/\Lambda} \Gamma \left( 0, \dfrac{\omega}{\Lambda} \right) \right].
\end{align}
Note that the first term diverges in the limit $\Lambda \to \infty$. The second term, however, vanishes in this limit. We see this by expanding it around the small dimensionless parameter $x \coloneqq \omega/\Lambda \ll 1$, which yields:
\begin{equation}
    e^{-x} \Ei \left( x \right) + e^{x} \Gamma \left( 0, x \right) \approx -2 x (-1 + \gamma_E + \log x) + \mathcal{O} (x^3) \approx 0,
\end{equation}
where $\gamma_E$ is the Euler-Mascheroni constant and we have used the limit $\lim_{x \to 0^+} x \log x = 0$. Consequently, in the limit $\Lambda \to \infty$, the boundary value becomes
\begin{equation}
    W_1^+ (\omega) = \dfrac{2 \gamma \Lambda}{\pi} + i \mathcal{J}_{\rm R} (\omega),
\end{equation}
allowing us to evaluate $W_0^+ (\omega)$ using Eq.~\eqref{WoW1} as
\begin{equation}\label{W0+final}
    W_0^+ (\omega) = \dfrac{g^2}{\Omega_{\rm R}^2 - \omega^2 - \frac{2 \gamma \Lambda}{\pi} - i \mathcal{J}_{\rm R} (\omega)} = \dfrac{g^2 \left( \Omega_{\rm R}^2 - \omega^2 - \frac{2 \gamma \Lambda}{\pi} + i \mathcal{J}_{\rm R} (\omega) \right)}{\left[ \Omega_{\rm R}^2 - \omega^2 - \frac{2 \gamma \Lambda}{\pi} \right]^2 + \mathcal{J}_{\rm R}^2 (\omega)}.
\end{equation}
To extend the relation between \(W_0(z)\) and \(W_1(z)\) to the corresponding boundary values \(W_0^+(\omega)\) and \(W_1^+(\omega)\), the denominator in Eq.~\eqref{W0+final} must not vanish. Using Eq.~\eqref{eqC18}, the denominator can be written as
\begin{equation}
\Omega_{\rm R}^2-\omega^2-W_1^+(\omega) = \Omega_{\rm R}^2-\omega^2-\Delta_1(\omega)-i\mathcal{J}_{\rm R}(\omega).
\end{equation}
For $\omega>0$, one has $\mathcal{J}_{\rm R}(\omega)>0$, meaning that the denominator cannot vanish. At $\omega=0$, since $\mathcal{J}_{\rm R}(0)=0$, the denominator reduces to
\begin{equation}
\Omega_{\rm R}^2-\Delta_1(0)
=
\Omega_{\rm R}^2-\frac{2\gamma\Lambda}{\pi},
\end{equation}
which corresponds precisely to the squared physical RC frequency in Eq.~\eqref{physicalRCfrequency}. Thus, the extension to the boundary values \(W_0^+(\omega)\) and \(W_1^+(\omega)\) is well-defined provided that the physical RC frequency does not vanish. 

Finally, taking the imaginary part $\mathcal{J} (\omega) = \Im W_0^+ (\omega)$ correctly reproduces Eq.~\eqref{SDlorentzian} as
\begin{equation}
    \mathcal{J} (\omega) = \dfrac{\gamma g^2 \omega}{\left[ \Omega_{\rm R}^2 - \omega^2 - \frac{2 \gamma \Lambda}{\pi} \right]^2 + \gamma^2 \omega^2},
\end{equation}
which correctly identifies the shift $\delta \Omega_{\rm R}^2 = \frac{2 \gamma \Lambda}{\pi}$. Indeed, evaluating Eq.~\eqref{deltaOmegaRC} yields
\begin{equation}\label{finalSDlorentzian}
    \delta \Omega_{\rm R}^2 = \sum_{\alpha \neq \rm R} \dfrac{C_\alpha^2}{\Omega_\alpha^2} = \dfrac{2}{\pi} \int_0^\infty \dd{\omega} \dfrac{\mathcal{J}_{\rm R} (\omega)}{\omega} = \dfrac{2}{\pi} \int_0^\infty \dd{\omega} \dfrac{\gamma \omega e^{-\omega/\Lambda}}{\omega} = \dfrac{2 \gamma}{\pi} \int_0^\infty \dd{\omega} e^{-\omega/\Lambda} = \dfrac{2 \gamma \Lambda}{\pi}.
\end{equation}
This result is consistent with Ref.~\cite{Iles_PRA2014}, where the effective spectral density $\mathcal{J} (\omega)$ explicitly depends on the physical reaction coordinate frequency rather than the bare frequency $\Omega_{\rm R}$.
\section{Continuously Monitoring a Single Mode of a Multi-Mode Gaussian System via a Gaussian Measurement}\label{app:Gaussian_meas}
Let us consider a multi-mode Gaussian system with covariance matrix $\sigma=\left(\begin{smallmatrix}
    \sigma_{\rm R} & C \\ C^T & \sigma_{\rm S}
\end{smallmatrix}\right)$ and displacement vector $d=\left[\begin{smallmatrix}
     d_{\rm R} \\ d_{\rm S}
\end{smallmatrix}\right]$. Here, the $\rm R$ block can be a multi-mode Gaussian system itself, representing the RCs extracted from the bath, while the S mode represents a single mode---the original system. We are interested in continuous monitoring of the system mode, while the S--RC unit is also in contact with a (residual) bath. As we are dealing with an overall Gaussian system, the assumption of divisibility of the dissipation and measurement maps---i.e., Eq.~\eqref{eq:cont_mon_divisible_evolution}---can be cast as divisibility of the conditional evolution of the first and second moments. We are interested in the infinitesimal evolution of the S--RC subject to both the dissipation and the measurement. The dissipative dynamics due to the residual bath is a standard Gaussian channel acting as
\begin{align}
    \sigma & \mapsto \sigma + \delta t [A_1(\theta) \sigma + \sigma A_1^T + D(\theta)]\eqqcolon  \bar \sigma  = \left[\begin{smallmatrix}
    \bar\sigma_{\rm R} & \bar C \\ \bar C^T & \bar \sigma_{\rm S}
\end{smallmatrix}\right],\nonumber\\
    d & \mapsto d + \delta t A_1(\theta) d \eqqcolon \bar d = \left[\begin{smallmatrix}
     \bar d_{\rm R} \\ \bar d_{\rm S},\end{smallmatrix}\right]
\end{align}
where the diffusion and drift matrices $A_1(\theta)$ and $D_1(\theta)$ depend on the S--RC Hamiltonian, as well as their dissipation to the residual bath. We leave them as general as possible here, but in Appendix~\ref{app:GKLSME} we provide a specific derivation for the example presented in the main text.

As already mentioned, the system also dissipates (leaks) into a secondary environment, where the measurement apparatus (detector) is placed, while the RC modes are not directly dissipating to this environment. Without considering the measurement record, the entire S--RC unit undergoes another unconditional open-system dynamics, which is another Gaussian channel
\begin{align}\label{eq:SRC_diss}
    \left[\begin{smallmatrix}
        \bar \sigma_{\rm R} & \bar C\\ \bar C^T & \bar \sigma_{\rm S} 
    \end{smallmatrix}\right] & \mapsto \left(\mathds{1}_{\rm R}\oplus  e^{-\lambda \delta t/2} \mathds{1}_{\rm S}\right)\left[\begin{smallmatrix}
        \bar \sigma_{\rm R} & \bar C\\ \bar C^T & \bar \sigma_{\rm S} 
    \end{smallmatrix}\right]\left(\mathds{1}_{\rm R}\oplus  e^{-\lambda \delta t/2} \mathds{1}_{\rm S}\right) + (1-e^{-\lambda \delta t})\left[\begin{smallmatrix}
        \mathds{O} & \mathds{O}\\ \mathds{O} & \sigma_{{\rm env}} 
    \end{smallmatrix}\right],\nonumber\\
    \left[\begin{smallmatrix}
        \bar d_{\rm R} \\ \bar d_{\rm S} 
    \end{smallmatrix}\right] & \mapsto \left(\mathds{1}_{\rm R}\oplus  e^{-\lambda \delta t/2} \mathds{1}_{\rm S}\right)\left[\begin{smallmatrix}
        \bar d_{\rm R} \\ \bar d_{\rm S} 
    \end{smallmatrix}\right].
\end{align}
Upon substituting for $\bar \sigma$ and $\bar d$, and keeping the terms up to linear in $\delta t$, this takes us back to Eqs.~\eqref{eq:Gaussian_map} of the main text. Note that, while the measurement-induced dynamics does not affect $\sigma_{\rm R}$ directly, it naturally leads to a decay of its correlations with $\sigma_{\rm S}$, captured by $C$. Furthermore, $\sigma_{\rm env}$ denotes the covariance matrix of the environment state, and will also be the steady state of the system---if this dynamics was the only one affecting the S--RC. For a thermal environment, $\sigma_{\rm env} = \nu \mathds{1} / 2$, with $\nu = \coth(\omega/2T)$. Without loss of generality, we take $d_{\rm env} = \mathds{O}$, but we set the environment to be at an arbitrary temperature, although it is usually chosen to be at $T=0$. The constant $\lambda > 0$ can be seen as the dissipation (leak) rate into the environment, also interpretable as the measurement strength in what follows.

To take the measurement record into account, we need to keep track of the environment mode, which is what enters the detector (see Fig.~\ref{fig:RC} in the main text). To this aim, note that the dissipative S--RC dynamics in Eqs.~\eqref{eq:SRC_diss} is the marginal channel of a partial swap between the system and an environment mode. The partial swap, in turn, is achieved by simply overlapping the system and $\sigma_{\rm env}$ on a beam splitter with transmissivity $\tau = e^{-\lambda \delta t}$---that is $S_{\rm BS}(\tau) = \left[\begin{smallmatrix}
    \sqrt{\tau} \mathds{1} & -\sqrt{1-\tau} \mathds{1} \\
    \sqrt{1-\tau} \mathds{1} & \sqrt{\tau} \mathds{1}
\end{smallmatrix}\right]$---to get
\begin{align}
    \left[ \begin{smallmatrix}
        \bar \sigma & \mathds{O}\\
        \mathds{O} & \sigma_{\rm env}
    \end{smallmatrix} \right]
    & \mapsto
    (\mathds{1}_{\rm R}\oplus S_{\rm BS}(\tau))\left[ \begin{smallmatrix}
        \bar \sigma & \mathds{O}\\
        \mathds{O} & \sigma_{\rm env}
    \end{smallmatrix} \right] (\mathds{1}_{\rm R}\oplus S^T_{\rm BS}(\tau))
    \nonumber\\
    & = \left[ \begin{smallmatrix}
        \bar \sigma_{\rm R} & \sqrt{e^{-\lambda \delta t}}\bar C & \sqrt{1-e^{-\lambda \delta t}}\bar C\\
        \sqrt{e^{-\lambda \delta t}}\bar C^T & e^{-\lambda \delta t}\bar \sigma_{\rm S} + (1-e^{-\lambda \delta t})\sigma_{\rm env} & \sqrt{e^{-\lambda \delta t}(1-e^{-\lambda \delta t})}(\bar \sigma_{\rm S} - \sigma_{\rm env}) \\ 
        \sqrt{1-e^{-\lambda \delta t}}\bar C^T & \sqrt{e^{-\lambda \delta t}(1-e^{-\lambda \delta t})}(\bar \sigma_{\rm S} - \sigma_{\rm env}) & e^{-\lambda \delta t}\sigma_{\rm env} + (1-e^{-\lambda \delta t})\bar \sigma_{\rm S}
    \end{smallmatrix} \right],\nonumber\\
    \left[ \begin{smallmatrix}
        \bar d_{\rm R}\\\bar d_{\rm S}\\
        \mathds{O}
    \end{smallmatrix} \right]%
    & \mapsto  \left(\mathds{1}_{\rm R} \oplus S_{\rm BS}(\tau) \right) \left[ \begin{smallmatrix}
        \bar d_{\rm R}\\\bar d_{\rm S}\\
        \mathds{O}
    \end{smallmatrix} \right]
    =
    \left[ \begin{smallmatrix}
        \bar d_{\rm R} \\ e^{-\lambda \delta t/2}\bar d_{\rm S}\\
        \sqrt{1-e^{-\lambda \delta t/2}} \bar d_{\rm S} 
    \end{smallmatrix} \right].
\end{align}
In this appendix, purely for convenience, we have taken the convention of using the symplectic phase space as ${\rm R}\oplus {\rm S} \oplus {\rm env}$, thus swapping the reaction coordinate and the system with respect to the main text. By tracing out the environment mode, one recovers the unconditional master equation for the S--RC. However, in the continuous monitoring setting we would like to keep this mode, since it enters the measurement apparatus, whose outcome is later used to update the conditional state of the system. At the level of infinitesimal evolutions---which is what we need for continuous monitoring---the above transformations expanded to leading order in $\lambda \delta t\ll 1$ read
\begin{align}
    \delta \left[ \begin{smallmatrix}
        \bar \sigma & \mathds{O}\\
        \mathds{O} & \sigma_{\rm env}
    \end{smallmatrix} \right] & = \lambda \delta t \left( \begin{smallmatrix}
            \mathds{O} & -\frac{\bar C}{2} & \frac{{\bar C}}{\sqrt{\lambda \delta t}}  \\
            -\frac{\bar C^T}{2} & \sigma_{\rm env} - \bar \sigma_{\rm S}& \frac{\bar \sigma_{\rm S}-\sigma_{\rm env}}{\sqrt{\lambda \delta t}} \\
            \frac{{\bar C^T}}{\sqrt{\lambda \delta t}}  & \frac{\bar \sigma_{\rm S}-\sigma_{\rm env}}{\sqrt{\lambda \delta t}} & \bar\sigma_{\rm S} - \sigma_{\rm env}
        \end{smallmatrix} \right)
        \nonumber\\
        & = \delta t \left[ (A_2 \oplus \mathds{O}) \left[ \begin{smallmatrix}
        \bar \sigma & \mathds{O}\\
        \mathds{O} & \sigma_{\rm env}
    \end{smallmatrix} \right] +  \left[ \begin{smallmatrix}
        \bar \sigma & \mathds{O}\\
        \mathds{O} & \sigma_{\rm env}
    \end{smallmatrix} \right] (A_2^T \oplus \mathds{O}) +  D_2 \oplus \mathds{O}\right] + \lambda \delta t \left( \begin{smallmatrix}
            \mathds{O} & \mathds{O} & \frac{{\bar C}}{\sqrt{\lambda \delta t}}  \\
            \mathds{O} & \mathds{O}& \frac{\bar \sigma_{\rm S}-\sigma_{\rm env}}{\sqrt{\lambda \delta t}} \\
            \frac{{ \bar C^T}}{\sqrt{\lambda \delta t}}  & \frac{\bar \sigma_{\rm S}-\sigma_{\rm env}}{\sqrt{\lambda \delta t}} & \bar \sigma_{\rm S} - \sigma_{\rm env}
        \end{smallmatrix} \right) 
        \nonumber\\
        \delta \left[ \begin{smallmatrix}
        \bar d_{\rm R}\\
        \bar d_{\rm S}\\
        \mathds{O} 
    \end{smallmatrix} \right]  & = \lambda \delta t \left[ \begin{smallmatrix}
        \mathds{O}\\
        -\frac{\bar d_{\rm S}}{2}\\
         \frac{\bar d_{\rm S}}{\sqrt{\lambda \delta t}} 
    \end{smallmatrix} \right] = \delta t(A_2\oplus \mathds{O}) \left[ \begin{smallmatrix}
        \bar d_{\rm R}\\
        \bar d_{\rm S}\\
        \mathds{O} 
    \end{smallmatrix} \right] + \sqrt{\lambda \delta t}\left[ \begin{smallmatrix}
        \mathds{O}\\
        \mathds{O}\\
        \bar d_{\rm S} 
    \end{smallmatrix} \right],
\end{align}
where we defined $A_2 = \mathds{O}_{\rm R} \oplus (-\lambda/2)\mathds{1}_{\rm S}$ and $D_2 = \mathds{O}_{\rm R} \oplus \lambda \sigma_{\rm env}$ acting on reaction coordinate and system unit. From here, one can simply use the output state of the environment as the input to the detector, where Gaussian measurements with covariance matrix $\sigma^M$ are performed, with the outcome distributed according to a normal distribution with mean $\xi = \sqrt{\lambda \delta t} \bar d_{\rm S}$ and variance $V = \sigma^M  + \sigma_{\rm env} + \lambda \delta t (\bar \sigma_{\rm S} - \sigma_{\rm env})$. Conditioned on the observed outcome, one needs to update the S--RC unit covariance matrix and displacement vector. As we discussed in the main text, upon consecutive repetition of this process and collecting many outcomes, this leads to Eqs.~\eqref{eq:Wiener_update_General} (see also Refs.~\cite{brask2021gaussian,serafini2023quantum,RevModPhys.84.621}).
\section{System Dynamics: Exact Solution vs. RC Master Equation}\label{sec:app_Exact_and_GKLS}
In this appendix, we present three alternative descriptions used throughout the manuscript to solve the dynamics of the model in Eq.~\eqref{eq:total_ham_original}. We first provide an exact solution based on the quantum Langevin equation. We then derive a global GKLS master equation for the system--RC unit and, finally, introduce the GKLS master equation for the system only, which serves as a baseline for comparison.

\subsection{Exact Dynamics via the Generalized Langevin Equation}\label{app:exactsolution}
We begin by exactly solving the dynamics of the system by tracing out the bath degrees of freedom. This yields a generalized quantum Langevin equation with a non-local memory kernel, from which both the transient dynamics and the steady state are obtained. This exact solution serves as a benchmark for the unconditional reduced dynamics of the system oscillator. Assuming an initially factorized state
\begin{equation}
    \rho(0)=\rho_{\rm S}(0) \otimes \rho_{\rm B},
\end{equation}
where $\rho_{\rm B}$ is a thermal state of the bath at inverse temperature $\beta=1/T$ and $\rho_{\rm S}(0)$ is an arbitrary Gaussian state of the system oscillator, we compute the Heisenberg equations of motion for the original Hamiltonian using Eqs.~\eqref{eq:A20}-\eqref{eq:A21}. One can then obtain a formal solution for each bath mode using standard differential equation techniques. The fundamental solutions to the homogeneous equation correspond to a free harmonic oscillator, namely, $\{ x_j^{(1)} (t), x_j^{(2)} (t) \} = \{ \cos (\omega_j t), \sin (\omega_j t) \}$. The particular solution is given by
\begin{equation}
    x_j^p (t) = u_j^{(1)}(t) x_j^{(1)}(t) + u_j^{(2)}(t) x_j^{(2)}(t),
\end{equation}
where the time-dependent coefficients $u_j^{(1)} (t)$ and $u_j^{(2)} (t)$ are defined according to the coupled equations
\begin{equation}
    \dot{u}_j^{(1)} = \dfrac{1}{W[x_j^{(1)} (t),x_j^{(2)} (t)]} \det \begin{pmatrix}
        0 & x_j^{(2)} (t) \\
        c_j x_\mathrm{S} (t) & \dot{x}_j^{(2)} (t)
    \end{pmatrix}, \quad \dot{u}_j^{(2)} = \dfrac{1}{W[x_j^{(1)} (t),x_j^{(2)} (t)]} \det \begin{pmatrix}
        x_j^{(1)} (t) & 0 \\
        \dot{x}_j^{(1)} (t) & c_j x_\mathrm{S} (t)
    \end{pmatrix},
\end{equation}
with $W[x_j^{(1)} (t),x_j^{(2)} (t)]$ denoting the Wronskian determinant. Integrating these expressions yields
\begin{align}
    u_j^{(1)} (t) &= - \dfrac{c_j}{\omega_j} \int_0^t \dd{t'} x_\mathrm{S} (t') \sin (\omega_j t'), \\
    u_j^{(2)} (t) &= \dfrac{c_j}{\omega_j} \int_0^t \dd{t'} x_\mathrm{S} (t') \cos (\omega_j t').
\end{align}
Combining these results with the initial conditions gives the final formal solution for the bath modes:
\begin{equation}\label{solutionODExk}
    x_j (t) = x_j (0) \cos (\omega_j t) + \dfrac{p_j (0)}{\omega_j} \sin (\omega_j t) + \dfrac{c_j}{\omega_j} \int_0^t \dd{t'} x_\mathrm{S} (t') \sin \left[ \omega_j (t-t') \right],
\end{equation}
where we used the trigonometric identity $\sin (A-B) = \sin A \cos B - \sin B \cos A$. Substituting this solution into Eq.~\eqref{eq:A20} gives
\begin{equation}
    \ddot{x}_\mathrm{S} (t) = - \Omega_\mathrm{S}^2 x_\mathrm{S} (t) - x_\mathrm{S} (t) \sum_j \dfrac{c_j^2}{\omega_j^2} + \sum_j c_j \left( x_j (0) \cos (\omega_j t) + \dfrac{p_j (0)}{\omega_j} \sin (\omega_j t) \right) + \sum_j \dfrac{c_j^2}{\omega_j} \int_0^t \dd{t'} x_\mathrm{S} (t') \sin \left[ \omega_j (t-t') \right].
\end{equation}
Using the definition of the bath spectral density in Eq.~\eqref{SDdiracdelta} and rearranging terms we get
\begin{equation}\label{eq123}
    \ddot{x}_\mathrm{S} (t) + (\Omega_\mathrm{S}^2 + \delta \Omega_\mathrm{S}^2) x_\mathrm{S} (t) - \dfrac{2}{\pi} \int_0^\infty \dd{\omega} \mathcal{J} (\omega) \int_0^t \dd{t'} x_\mathrm{S} (t') \sin \left[ \omega (t-t') \right] = \sum_j \tilde{B}_j(t)\eqqcolon {\tilde B}(t),
\end{equation}
where we have introduced the bath-noise operator
\begin{equation}\label{eqa8}
    \tilde{B}_j(t) \coloneqq c_j \left( x_j (0) \cos (\omega_j t) + \dfrac{p_j (0)}{\omega_j} \sin (\omega_j t) \right).
\end{equation}
By defining the dissipation kernel as
\begin{equation}
    \chi (t) \coloneqq \dfrac{2}{\pi} \int_0^\infty \dd{\omega'} \mathcal{J} (\omega') \sin \left( \omega' t \right),
\end{equation}
we obtain the standard generalized Langevin equation:
\begin{equation}
    \ddot{x}_\mathrm{S} (t)
    + (\Omega_\mathrm{S}^2 + \delta \Omega_\mathrm{S}^2)x_\mathrm{S}(t)
    - \int_0^t \dd t'\,\chi(t-t')x_\mathrm{S}(t')
    = \tilde B(t).
\end{equation}
Since the memory integral runs over $0\leq t'\leq t$, the argument $t-t'$ is strictly non-negative, and no explicit Heaviside factor is needed. The exact solution can then be written in terms of the Green function $G(t)$ as \cite{Lampo2017}
\begin{equation}\label{eqa11}
    x_\mathrm{S}(t)
    = x_\mathrm{S}(0)\dot G(t)
    + p_\mathrm{S}(0) G(t)
    + \int_0^t \dd\tau\, G(t-\tau)\tilde B(\tau),
\end{equation}
where $G (t)$ and $\dot{G} (t)$ are defined via the Laplace-space Green function $\mathcal{G} (s)$ according to
\begin{align}
    \mathcal{L} \{ G (t) \} \eqqcolon \mathcal G(s)
    &= \frac{1}{s^2+\Omega_\mathrm{S}^2+\delta\Omega_\mathrm{S}^2-\mathcal L\{\chi\}(s)}, \label{greenfunction} \\
    \mathcal{L} \{ \dot{G} (t) \} &= \frac{s}{s^2+\Omega_\mathrm{S}^2+\delta\Omega_\mathrm{S}^2-\mathcal L\{\chi\}(s)}. \label{greenfunctiondot}
\end{align}
To reproduce the correct initial conditions, Eq.~\eqref{eqa11} strictly requires $G (0) = 0$ and $\dot{G} (0) = 1$. Differentiating Eq.~\eqref{eqa11} provides the exact solution for the momentum quadrature:
\begin{equation}\label{eqpS}
    p_{\rm S} (t) = x_{\rm S} (0) \Ddot{G} (t) + p_{\rm S} (0) \dot{G} (t) + \int_0^t \dd{\tau} \dot{G} (t-\tau) \Tilde{B} (\tau).
\end{equation}
\subsubsection{Laplace-Space Green Functions}
To obtain an analytical expression for the Green function in $s$-space, we first redefine the parameters of the spectral density in Eq.~\eqref{SDlorentzian} using $g^2 = \alpha_1 \alpha_2 \gamma$ and $\Omega_{\rm R}^2 = \alpha_2 \gamma$, which gives
\begin{equation}\label{brownianSD}
    \mathcal{J} (\omega) = \dfrac{\gamma^2 \alpha_1 \alpha_2 \omega}{\gamma^2 \omega^2 + (\alpha_2 \gamma - \omega^2)^2}.
\end{equation}
We now find $\mathcal{L} \{ \chi \} (s)$ by evaluating the following integral:
\begin{equation}
    \mathcal{L} \{ \chi \} (s) = \dfrac{2}{\pi} \int_0^\infty \dd{\omega} \mathcal{J} (\omega) \dfrac{\omega}{\omega^2 + s^2}.
\end{equation}
Substituting the spectral density from Eq.~\eqref{brownianSD} into the integral yields
\begin{equation}
    \mathcal{L} \{ \chi \} (s) = \dfrac{2 \gamma^2 \alpha_1 \alpha_2}{\pi} \int_0^\infty \dd{\omega} \dfrac{1}{\gamma^2 \omega^2 + (\alpha_2 \gamma - \omega^2)^2} \dfrac{\omega^2}{\omega^2 + s^2}.
\end{equation}
To evaluate this integral using the residue theorem, we symmetrize it over the entire real line as follows:
\begin{equation}
    \mathcal{L} \{ \chi \} (s) = \dfrac{\gamma^2 \alpha_1 \alpha_2}{\pi} \int_{-\infty}^\infty \dd{\omega} \dfrac{1}{\gamma^2 \omega^2 + (\alpha_2 \gamma - \omega^2)^2} \dfrac{\omega^2}{\omega^2 + s^2} \eqqcolon \int_{-\infty}^\infty \dd{\omega} f(\omega),
\end{equation}
allowing the theorem to be applied over a closed contour $\mathcal{C}$ in the upper half-plane:
\begin{equation}
    \oint_\mathcal{C} \dd{z} f(z) = 2 \pi i \sum_j \Res (f, \omega_j),
\end{equation}
with $\Res (f, \omega_j)$ being the residue of the function $f (z)$ at pole $\omega_j$. The relevant poles of the function $f(\omega)$ in the upper half-plane are
\begin{align}
    \omega_1 &= i s, \\
    \omega_2^\pm &= \dfrac{i \gamma \pm \sqrt{4 \alpha_2 \gamma - \gamma^2}}{2},
\end{align}
with corresponding residues
\begin{align}
    \Res (f(\omega), \omega_1) &= \dfrac{i s}{2 \left[ (\alpha_2 \gamma + s^2)^2 - \gamma^2 s^2 \right]}, \\
    \Res (f(\omega), \omega_2^+) + \Res (f(\omega), \omega_2^-) &= - \dfrac{i (\alpha_2 \gamma + s^2)}{2 \gamma \left[ (\alpha_2 \gamma + s^2)^2 - \gamma^2 s^2 \right]}.
\end{align}
Summing these residues provides the evaluated integral for $\mathcal{L} \{ \chi \} (s)$:
\begin{equation}
    \mathcal{L} \{ \chi \} (s) = \dfrac{\gamma^2 \alpha_1 \alpha_2}{(\alpha_2 \gamma + s^2)^2 - \gamma^2 s^2} \left( \alpha_2 + \dfrac{s^2}{\gamma} - s \right) = \dfrac{\gamma \alpha_1 \alpha_2}{s^2 + s \gamma + \alpha_2 \gamma}.
\end{equation}
Substituting this result back into Eq.~\eqref{greenfunction}, we obtain the explicit Laplace-space Green function
\begin{equation}
    \mathcal{G} (s) = \dfrac{1}{s^2 + \Omega_{\rm S}^2 + \delta \Omega_{\rm S}^2 - \frac{\gamma \alpha_1 \alpha_2}{s^2 + s \gamma + \alpha_2 \gamma}} = \dfrac{s^2 + s \gamma + \alpha_2 \gamma}{(s^2 + \Omega_{\rm S}^2 + \delta \Omega_{\rm S}^2) (s^2 + s \gamma + \alpha_2 \gamma) - \gamma \alpha_1 \alpha_2}.
\end{equation}
To solve the time-domain dynamics, $G(t)$ is recovered by computing the inverse Laplace transform
\begin{equation}
    G (t) = \dfrac{1}{2 \pi i} \int_{\sigma - i \infty}^{\sigma + i \infty} \dd{s} e^{st} \mathcal{G} (s),
\end{equation}
where $\sigma \in \mathbb{R}$ ensures the contour path lies within the region of convergence of $\mathcal{G} (s)$. Because $\mathcal{G} (s)$ is a rational function of the form $\mathcal{G} (s) = B(s) / A(s)$, with $B(s) \coloneqq s^2 + s \gamma + \alpha_2 \gamma$ and $A (s) \coloneqq (s^2 + \Omega_{\rm S}^2 + \delta \Omega_{\rm S}^2) (s^2 + s \gamma + \alpha_2 \gamma) - \gamma \alpha_1 \alpha_2$, the inverse transform can be evaluated analytically. Defining $\{p_1, \dots, p_4\}$ as the simple roots of $A (s)$, the residues of $\mathcal{G} (s)$ are given by $c_i = \left[ (s-p_i) \mathcal{G} (s) \right]_{s = p_i}$. The inverse Laplace transform is then constructed directly from the sum of these residues as
\begin{equation}
    G (t) = \left( c_1 e^{p_1 t} + \dots + c_4 e^{p_4 t} \right) \Theta(t).
\end{equation}
Since the roots are simple for our parameter choices, finding $G(t)$ simply requires computing the roots and corresponding residues of $\mathcal{G} (s)$, which we perform numerically.

\subsubsection{System Covariances}
To characterize the unconditional Gaussian dynamics, we explicitly compute the symmetrized covariances $\ev{x_{\rm S}^2 (t)}$ and $\ev{p_{\rm S}^2 (t)}$. Squaring Eq.~\eqref{eqa11} and taking the expectation value separates the variance into system and bath contributions as follows:
\begin{equation}
    \ev{x_{\rm S}^2 (t)} = \ev{x_{\rm S}^2 (t)}^{\rm sys} + \ev{x_{\rm S}^2 (t)}^{\rm bath},
\end{equation}
with the respective contributions given by
\begin{align}
    \ev{x_{\rm S}^2 (t)}^{\rm sys} &= \dot{G} (t)^2 \ev{x_{\rm S}^2 (0)} + G (t)^2 \ev{p_{\rm S}^2 (0)} + G (t) \dot{G} (t) \ev{x_{\rm S} (0) p_{\rm S} (0) + p_{\rm S} (0) x_{\rm S} (0)}, \notag \\
    \ev{x_{\rm S}^2 (t)}^{\rm bath} &= \int_0^t \dd{\tau} \int_0^t \dd{\tau'} G (t-\tau) \nu(\tau - \tau') G (t-\tau'), 
\end{align}
where we used the fact that the initial thermal bath state implies $\ev{\Tilde{B} (\tau)} = 0$. The symmetrized noise kernel $\nu (\tau - \tau')$ is defined as
\begin{equation}
    \nu (\tau - \tau') \coloneqq \dfrac{1}{2} \ev{\acomm{\Tilde{B} (\tau)}{\Tilde{B} (\tau')}},
\end{equation}
which, using Eq.~\eqref{eqa8}, can be expressed directly in terms of the spectral density as
\begin{equation}\label{noisekernelexplicit}
    \nu (\tau - \tau') = \dfrac{1}{\pi} \int_0^\infty \dd{\omega} \mathcal{J} (\omega) \coth \left( \dfrac{\beta \omega}{2} \right) \cos \left[ \omega (\tau - \tau') \right].
\end{equation}
Similarly, squaring Eq.~\eqref{eqpS} and taking the expectation value gives the momentum variance contributions
\begin{align}
    \ev{p_{\rm S}^2 (t)}^{\rm sys} &= \Ddot{G} (t)^2 \ev{x_{\rm S}^2 (0)} + \dot{G} (t)^2 \ev{p_{\rm S}^2 (0)} + \dot{G} (t) \Ddot{G} (t) \ev{x_{\rm S} (0) p_{\rm S} (0) + p_{\rm S} (0) x_{\rm S} (0)}, \notag \\
    \ev{p_{\rm S}^2 (t)}^{\rm bath} &= \int_0^t \dd{\tau} \int_0^t \dd{\tau'} \dot{G} (t-\tau) \nu(\tau - \tau') \dot{G} (t-\tau').
\end{align}
Evaluating the noise kernel in Eq.~\eqref{noisekernelexplicit} again relies on the residue theorem, as detailed below. Finally, the off-diagonal terms of the symmetrized covariance matrix will read
\begin{align}
    \frac{1}{2}\ev{\{x_{\rm S}(t), p_{\rm S}(t)\}}^{\rm sys} &= \;  \dot{G}(t)\ddot{G}(t)\ev{x_{\rm S}^2(0)} +  G(t)\dot{G}(t)\ev{p_{\rm S}^2(0)} + \frac{1}{2}\left[ \dot{G}(t)^2 + G(t)\ddot{G}(t) \right] \ev{x_{\rm S} (0) p_{\rm S} (0) + p_{\rm S} (0) x_{\rm S} (0)}, \notag \\
    \frac{1}{2}\ev{\{x_{\rm S}(t), p_{\rm S}(t)\}}^{\rm bath} &= \int_0^t \dd{\tau} \int_0^t \dd{\tau'} G(t-\tau)\nu(\tau-\tau')\dot{G}(t-\tau').
\end{align}

\subsubsection{Evaluation of the Noise Kernel}
To evaluate the noise kernel in Eq.~\eqref{noisekernelexplicit} using the residue theorem, we rewrite it as
\begin{equation}
    \nu (t) = \dfrac{\gamma^2 \alpha_1 \alpha_2}{2 \pi} \int_{-\infty}^{\infty} \dd{\omega} \Re \left[ \dfrac{\omega}{\gamma^2 \omega^2 + (\alpha_2 \gamma - \omega^2)^2} \coth \left( \dfrac{\beta \omega}{2} \right) e^{i \omega t} \right] \eqqcolon \int_{-\infty}^\infty \dd{\omega} F(\omega).
\end{equation}
The function $F(\omega)$ possesses poles at
\begin{align}
    \omega_1^{(k)} &= i f_k, \quad f_k = \dfrac{2 \pi k}{\beta}, \quad k \in \mathbb{Z} \\
    \omega_2^\pm &= \dfrac{i \gamma \pm \sqrt{4 \alpha_2 \gamma - \gamma^2}}{2}.
\end{align}
Accounting for the infinite series of Matsubara poles, the residue sum yields the explicit time-dependent kernel
\begin{align}
    \nu (t) &= \dfrac{\gamma \alpha_1 \alpha_2}{2 \sqrt{4 \alpha_2 \gamma - \gamma^2}} \Re \left[ \coth \left( \dfrac{\beta \omega_2^+}{2} \right) e^{i \omega_2^+ t} - \coth \left( \dfrac{\beta \omega_2^-}{2} \right) e^{i \omega_2^- t} \right] - \dfrac{2 \gamma^2 \alpha_1 \alpha_2}{\beta} \sum_{k=1}^\infty \dfrac{f_k e^{-f_k t}}{(\alpha_2 \gamma + f_k^2)^2 - \gamma^2 f_k^2}. \notag
\end{align}

\subsection{Global GKLS Master Equation for the System--RC Unit}\label{app:GKLSME}
We now derive a global master equation for the combined system--reaction coordinate unit. Since the system and reaction coordinate interact strongly, we first diagonalize the augmented system Hamiltonian
\begin{equation}
    H_{\rm S-RC} = \dfrac{1}{2} (p_{\rm S}^2 + \Omega_1^2 x_{\rm S}^2) + \dfrac{1}{2} (P_{\rm R}^2 + \Omega_2^2 X_{\rm R}^2) - g x_{\rm S} X_{\rm R},
\end{equation}
where we have defined $\Omega_1^2 \coloneqq \Omega_{\rm S}^2 + \delta \Omega_{\rm S}^2$ and $\Omega_2^2 \coloneqq \Omega_{\rm R}^2$. Here, the renormalization shift is absorbed directly into $\Omega_1$ to simplify the subsequent matrix definitions. The normal mode frequencies of this Hamiltonian are given by
\begin{align}
    \Omega_{\pm}^2 = \frac{1}{2}\left(\Omega_1^2 + \Omega_2^2 \pm \sqrt{4g^2 + (\Omega_1^2 - \Omega_2^2)^2} \right),
\end{align}
with the corresponding free quadratures defined as
\begin{align}
    \begin{pmatrix}
        u_+ \\ u_-
    \end{pmatrix} 
    & = 
    \begin{pmatrix}
        \cos\theta & -\sin\theta\\ \sin\theta & \cos\theta
    \end{pmatrix} 
    \begin{pmatrix}
        x_{\rm S} \\ X_{\rm R}
    \end{pmatrix}, \quad \text{with} \quad \cos^2\theta = \frac{\Omega_1^2 - \Omega_2^2 + \sqrt{4g^2 + (\Omega_1^2 - \Omega_2^2)^2}}{2\sqrt{4g^2 + (\Omega_1^2 - \Omega_2^2)^2}}.
\end{align}
The four associated jump operators $\{b_+, b_-, b_+^{\dagger}, b_-^{\dagger}\}$ are then constructed as
\begin{align}
     \begin{pmatrix}
            b_+\\b_+^{\dagger}\\b_-\\b_-^{\dagger}
     \end{pmatrix}
    & = \dfrac{1}{\sqrt{2}} 
     \begin{pmatrix}
            \Omega_+^{1/2} & i\Omega_+^{-1/2} & 0 & 0\\
            \Omega_+^{1/2} & -i\Omega_+^{-1/2} & 0 & 0\\
            0 & 0 & \Omega_-^{1/2} & i\Omega_-^{-1/2}\\
            0 & 0 & \Omega_-^{1/2} & -i\Omega_-^{-1/2}
     \end{pmatrix}
     \begin{pmatrix}
            u_+\\{\dot u}_+\\u_-\\{\dot u}_-
     \end{pmatrix}
     \nonumber\\
     &=
     \dfrac{1}{\sqrt{2}}
     \begin{pmatrix}
            \Omega_+^{1/2} & i\Omega_+^{-1/2} & 0 & 0\\
            \Omega_+^{1/2} & -i\Omega_+^{-1/2} & 0 & 0\\
            0 & 0 & \Omega_-^{1/2} & i\Omega_-^{-1/2}\\
            0 & 0 & \Omega_-^{1/2} & -i\Omega_-^{-1/2}
     \end{pmatrix}
    \begin{pmatrix}
        \cos\theta & 0 & -\sin\theta & 0 \\
        0 & \cos\theta & 0 & -\sin\theta \\ 
        \sin\theta & 0 & \cos\theta & 0 \\
        0 & \sin\theta & 0 & \cos\theta 
    \end{pmatrix}
     \begin{pmatrix}
            x_{\rm S}\\p_{\rm S}\\X_{\rm R}\\P_{\rm R}
     \end{pmatrix}\nonumber\\
     & = 
     \dfrac{1}{\sqrt{2}}
     \begin{pmatrix}
            \Omega_+^{1/2} \cos\theta & i\Omega_+^{-1/2} \cos\theta & -\Omega_+^{1/2} \sin\theta & -i\Omega_+^{-1/2} \sin\theta  \\ 
            \Omega_+^{1/2} \cos\theta & -i\Omega_+^{-1/2} \cos\theta & -\Omega_+^{1/2} \sin\theta & i\Omega_+^{-1/2} \sin\theta \\ 
            \Omega_-^{1/2} \sin\theta & i\Omega_-^{-1/2} \sin\theta  &\Omega_-^{1/2} \cos\theta & i\Omega_-^{-1/2} \cos\theta  \\ 
            \Omega_-^{1/2} \sin\theta & -i\Omega_-^{-1/2} \sin\theta  &\Omega_-^{1/2} \cos\theta & -i\Omega_-^{-1/2} \cos\theta
     \end{pmatrix}
     \begin{pmatrix}
            x_{\rm S}\\p_{\rm S}\\X_{\rm R}\\P_{\rm R}
     \end{pmatrix}. \label{transformationglobal}
\end{align}
To derive the open-system dynamics, we partition the total Hamiltonian into the system--RC unit, the interaction, and the residual bath as follows:
\begin{equation}
    H = H_{\rm S-RC} - \underbrace{X_{\rm R} \sum_{\alpha \neq \rm R} C_\alpha X_\alpha}_{H_{\rm I}} + \underbrace{\dfrac{1}{2} \sum_{\alpha \neq \rm R} (P_\alpha^2 + \Omega_\alpha^2 X_\alpha^2)}_{H_{\rm RB}}.
\end{equation}
Moving into the interaction picture, we write the interaction Hamiltonian as
\begin{equation}
\Tilde{H}_\mathrm{I} (t) = e^{i (H_\mathrm{S-RC} + H_\mathrm{RB}) t} H_\mathrm{I} e^{-i (H_\mathrm{S-RC} + H_\mathrm{RB}) t}.
\end{equation}
The dynamics of the reduced density matrix in the interaction picture is governed by the following equation:
\begin{equation}
    \dot{\tilde{\rho}} = -i \Tr_{\rm B} \comm{\Tilde{H}_\mathrm{I} (t)}{\Tilde{\rho} (0)} - \int_0^t \dd{s} \Tr_{\rm B} \comm{\Tilde{H}_\mathrm{I} (t)}{\comm{\Tilde{H}_\mathrm{I} (s)}{\Tilde{\rho} (s)}}.
\end{equation}
Since the bath remains in a thermal state, the first term vanishes, simplifying the previous equation to
\begin{align}
    \dot{\tilde{\rho}} &= - \int_0^t \dd{s} \Tr_{\rm B} \comm{\Tilde{H}_\mathrm{I} (t)}{\comm{\Tilde{H}_\mathrm{I} (t - s)}{\Tilde{\rho}_\mathrm{S-RC} (t - s) \otimes \Tilde{\rho}_{\rm B}}} \\
    &= - \int_0^t \dd{s} \Tr_{\rm B} \left[ \Tilde{H}_\mathrm{I} (t) \Tilde{H}_\mathrm{I} (t - s) \left( \Tilde{\rho}_\mathrm{S-RC} (t - s) \otimes \Tilde{\rho}_{\rm B} \right) - \Tilde{H}_\mathrm{I} (t - s) \left( \Tilde{\rho}_\mathrm{S-RC} (t - s) \otimes \Tilde{\rho}_{\rm B} \right) \Tilde{H}_\mathrm{I} (t) + \mathrm{h.c.} \right] \\
    &= - \int_0^t \dd{s} \left( \Tilde{A} (t) \Tilde{A} (t-s) \Tilde{\rho}_\mathrm{S-RC} (t-s) - \Tilde{A} (t-s) \Tilde{\rho}_\mathrm{S-RC} (t-s) \Tilde{A} (t) \right) \Tr_{\rm B} \left( \Tilde{B} (t) \Tilde{B} (t-s) \Tilde{\rho}_{\rm B} \right) + \mathrm{h.c.},
\end{align}
where we changed the integration variable $s \leftrightarrow t-s$ and defined the operators
\begin{align}
    A = X_\mathrm{R} &\coloneqq \dfrac{b_- + b_-^\dagger}{\sqrt{2 \Omega_-}} \cos \theta - \dfrac{b_+ + b_+^\dagger}{\sqrt{2 \Omega_+}} \sin \theta \Rightarrow \Tilde{A} (t) = \dfrac{b_- e^{-i \Omega_- t} + b_-^\dagger e^{i \Omega_- t}}{\sqrt{2 \Omega_-}} \cos \theta - \dfrac{b_+ e^{-i \Omega_+ t} + b_+^\dagger e^{i \Omega_+ t}}{\sqrt{2 \Omega_+}} \sin \theta, \label{systemoperatorsIPglobal} \\
    B &\coloneqq - \sum_{\alpha \neq \rm R} \dfrac{C_\alpha}{\sqrt{2 \Omega_\alpha}} (b_\alpha + b_\alpha^\dagger) \Rightarrow \Tilde{B} (t) = - \sum_{\alpha \neq \rm R} \dfrac{C_\alpha}{\sqrt{2 \Omega_\alpha}} (e^{-i \Omega_\alpha t} b_\alpha + e^{i \Omega_\alpha t} b_\alpha^\dagger). \label{Btilde}
\end{align}
The bath correlation function evaluated over the thermal state is given by
\begin{equation}\label{bathcorrfuncglobal}
    \Tr_{\rm B} \left( \Tilde{B} (t) \Tilde{B} (t-s) \Tilde{\rho}_{\rm B} \right) = \dfrac{1}{2} \sum_{\alpha \neq \rm R} \dfrac{C_\alpha^2}{\Omega_\alpha} \left[ e^{-i \Omega_\alpha s} \left( \mathcal{N} (\Omega_\alpha, T) + 1 \right) + e^{i \Omega_\alpha s} \mathcal{N} (\Omega_\alpha, T) \right],
\end{equation}
where $\mathcal{N} (\Omega_\alpha, T)$ is the bosonic occupation number at temperature $T$ and frequency $\Omega_\alpha$. Eq.~\eqref{systemoperatorsIPglobal} indicates that the RC coupling operator oscillates at frequencies $\pm \Omega_\pm$. Under the Born--Markov and secular approximations, rapidly rotating cross terms between different normal modes are discarded, and only terms with matching Bohr frequencies are retained. Neglecting the imaginary Lamb shift terms\footnote{Here, $\Omega_{\rm R}$ is the physical RC frequency defined in Eq.~\eqref{physicalRCfrequency}. For a sufficiently large residual-bath cutoff, the coherent effect of the potential renormalization is approximately canceled by the
Lamb shift; see Ref.~\cite{Correa2025potential} for further discussion. The residual contribution was verified numerically to be negligible in the regime considered.} and going back to the Schrödinger picture, the resulting master equation assumes the standard GKLS form
\begin{equation}\label{gklsmegeneral}
    \dot{\rho} = -i \comm{H_{\rm S-RC}}{\rho} + \sum_\mu \left( L_\mu \rho L_\mu^\dagger - \dfrac{1}{2} \acomm{L_\mu^\dagger L_\mu}{\rho} \right),
\end{equation}
with the following set of Lindblad jump operators
\begin{align}\label{lindbladoperators}
    L_1 &= \left[ \dfrac{\sin^2 \theta}{\Omega_+} \mathcal{J}_\mathrm{R} (\Omega_+) \left( \mathcal{N} (\Omega_+, T) + 1 \right) \right]^{1/2} b_+, \\
    L_2 &= \left[ \dfrac{\sin^2 \theta}{\Omega_+} \mathcal{J}_\mathrm{R} (\Omega_+) \mathcal{N} (\Omega_+, T) \right]^{1/2} b_+^\dagger, \\
    L_3 &= \left[ \dfrac{\cos^2 \theta}{\Omega_-} \mathcal{J}_\mathrm{R} (\Omega_-) \left( \mathcal{N} (\Omega_-, T) + 1 \right) \right]^{1/2} b_-, \\
    L_4 &= \left[ \dfrac{\cos^2 \theta}{\Omega_-} \mathcal{J}_\mathrm{R} (\Omega_-) \mathcal{N} (\Omega_-, T) \right]^{1/2} b_-^\dagger.
      \label{lindbladoperators4}
\end{align}
For a Lindbladian master equation of this form, the dynamics of the first and second moments is governed by Eq.~\eqref{eq:Gaussian_map}. We construct the drift matrix $A$ and diffusion matrix $D$ according to
\begin{align}
  A &:= \Omega \left[ G + \Im (C^\dagger C) \right] \\
  D &:= \Omega \Re (C^\dagger C) \Omega^{\rm T}
\end{align}
where the symplectic matrix $\Omega$ is defined in the main text and $G$ captures the unitary dynamics
\begin{equation}
G \coloneqq
\begin{pmatrix}
\Omega_1^2 & 0 & -g & 0\\
0 & 1 & 0 & 0\\
-g & 0 & \Omega_2^2 & 0 \\
0 & 0 & 0 & 1
\end{pmatrix}.
\end{equation}
The matrix $C \coloneqq (c_1^\mathrm{T}; c_2^\mathrm{T}; c_3^\mathrm{T}; c_4^\mathrm{T})^\mathrm{T}$ is formed from the row vectors $c_\mu$, which define the Lindblad operators via $L_\mu = c_\mu^\mathrm{T} R$. Using the operators in Eqs.~\eqref{lindbladoperators}--\eqref{lindbladoperators4}, we construct $C$ by multiplying the transformation matrix by the corresponding weights
\begin{align}
    C
     & =
     \begin{pmatrix}
           \left[ \frac{\mathcal{J}_\mathrm{R}(\Omega_+) (\mathcal{N}(\Omega_+, T) + 1)}{2\Omega_+} \right]^{1/2} \sin\theta \\
          \left[ \frac{\mathcal{J}_\mathrm{R}(\Omega_+)\mathcal{N}(\Omega_+, T) }{2\Omega_+} \right]^{1/2} \sin\theta \\
           \left[ \frac{\mathcal{J}_\mathrm{R}(\Omega_-)(\mathcal{N}(\Omega_-, T) + 1)}{2\Omega_-} \right]^{1/2} \cos\theta \\
           \left[ \frac{\mathcal{J}_\mathrm{R}(\Omega_-)\mathcal{N}(\Omega_-, T)}{2\Omega_-} \right]^{1/2} \cos\theta 
     \end{pmatrix}
    \bullet
     \begin{pmatrix}
            \Omega_+^{1/2} \cos\theta & i\Omega_+^{-1/2} \cos\theta & -\Omega_+^{1/2} \sin\theta & -i\Omega_+^{-1/2} \sin\theta  \\ 
            \Omega_+^{1/2} \cos\theta & -i\Omega_+^{-1/2} \cos\theta & -\Omega_+^{1/2} \sin\theta & i\Omega_+^{-1/2} \sin\theta \\ 
            \Omega_-^{1/2} \sin\theta & i\Omega_-^{-1/2} \sin\theta  &\Omega_-^{1/2} \cos\theta & i\Omega_-^{-1/2} \cos\theta  \\ 
            \Omega_-^{1/2} \sin\theta & -i\Omega_-^{-1/2} \sin\theta  &\Omega_-^{1/2} \cos\theta & -i\Omega_-^{-1/2} \cos\theta
     \end{pmatrix},
\end{align}
where $v \bullet M \coloneqq {\rm diag}(v) M$ denotes row-wise multiplication. From this, the symmetric diffusion matrix $D$ is evaluated as
\begin{equation}
D =
\begin{pmatrix}
d_{11} & 0 & d_{13} & 0\\[0.8ex]
0 & d_{22} & 0 & d_{24} \\[0.8ex]
d_{31} & 0 & d_{33} & 0\\[0.8ex]
0 & d_{42} & 0 & d_{44}
\end{pmatrix}
\end{equation}
with elements given by
\begin{align}
d_{11} &= \dfrac{\sin^2(2\theta)}{8 \Omega_+^2\,\Omega_-^2}
\Big[\mathcal{J}_{\rm R} (\Omega_-) \Omega_+^2 (2\mathcal{N} (\Omega_-, T)+1)+\mathcal{J}_{\rm R} (\Omega_+) \Omega_-^2(2\mathcal{N} (\Omega_+, T) +1)\Big], \\
d_{22} &= \dfrac{\sin^2(2\theta)}{8}\Big[\mathcal{J}_{\rm R} (\Omega_-)(2\mathcal{N} (\Omega_-, T)+1)+\mathcal{J}_{\rm R} (\Omega_+)(2\mathcal{N} (\Omega_+, T) +1)\Big], \\
d_{33} &= \dfrac{\mathcal{J}_{\rm R} (\Omega_-) \Omega_+^2(2\mathcal{N} (\Omega_-, T)+1)\cos^4\theta
      +\mathcal{J}_{\rm R} (\Omega_+) \Omega_-^2(2 \mathcal{N} (\Omega_+, T) +1)\sin^4\theta}{2 \Omega_+^2\,\Omega_-^2}, \\
d_{44} &= \dfrac{1}{2}\Big[\mathcal{J}_{\rm R} (\Omega_-)(2 \mathcal{N} (\Omega_-, T)+1)\cos^4\theta
               +\mathcal{J}_{\rm R} (\Omega_+)(2 \mathcal{N} (\Omega_+, T) +1)\sin^4\theta\Big], \\
d_{24} = d_{42}
&= \frac{\sin(2\theta)}{4}\Big[\mathcal{J}_{\rm R} (\Omega_-)(2 \mathcal{N} (\Omega_-, T) +1)\cos^2\theta
- \mathcal{J}_{\rm R} (\Omega_+)(2 \mathcal{N} (\Omega_+, T) +1)\sin^2\theta\Big],\\[0.6ex]
d_{13} = d_{31}
&= \frac{\sin(2\theta)}{4 \Omega_+^2\,\Omega_-^2}\Big[\mathcal{J}_{\rm R} (\Omega_-) \Omega_+^2(2 \mathcal{N} (\Omega_-, T) +1)\cos^2\theta
- \mathcal{J}_{\rm R} (\Omega_+) \Omega_-^2(2 \mathcal{N} (\Omega_+, T) +1)\sin^2\theta\Big].
\end{align}
Similarly, the drift matrix $A$ evaluates to
\begin{equation}
A =
\begin{pmatrix}
a_{11} & 1 & a_{13} & 0\\[0.8ex]
-\Omega_1^2
& a_{22}& g & a_{24}\\[0.8ex]
a_{31} & 0 & a_{33} & 1\\[0.8ex]
g & a_{42} & -\Omega_2^2 & a_{44}
\end{pmatrix}
\end{equation}
where
\begin{align}
a_{11} = a_{22} &= -\dfrac{\sin^2(2\theta)}{8\,\Omega_+\Omega_-}\Big(\mathcal{J}_{\rm R} (\Omega_-) \Omega_+ + \mathcal{J}_{\rm R} (\Omega_+) \Omega_-\Big), \\[1.0ex]
a_{33} = a_{44}
&= -\frac{1}{2\,\Omega_+\Omega_-}
\Big[
\mathcal{J}_{\rm R} (\Omega_-) \Omega_+ \cos^4\theta
+ \mathcal{J}_{\rm R} (\Omega_+) \Omega_- \sin^4\theta
\Big], \\[1.0ex]
a_{13} = a_{31} = a_{24} = a_{42}
&= -\frac{\sin(2\theta)}{4\,\Omega_+\Omega_-}\Big(\mathcal{J}_{\rm R} (\Omega_-) \Omega_+ \cos^2 \theta - \mathcal{J}_{\rm R} (\Omega_+) \Omega_-\sin^2\theta\Big).
\end{align}
As a consistency check, we can verify the signs of the drift components by taking the limit of zero dissipation, $\mathcal{J}_{\rm R} (\Omega_\pm) \to 0$. In this limit, the dissipative contributions vanish, and the drift matrix reduces purely to the unitary generator
\begin{equation}
    A \to 
    \begin{pmatrix}
        0 & 1 & 0 & 0 \\
        -\Omega_1^2 & 0 & g & 0 \\
        0 & 0 & 0 & 1 \\
        g & 0 & -\Omega_2^2 & 0
    \end{pmatrix}.
\end{equation}
This correctly recovers the classical Hamilton equations of motion for the isolated system-RC Hamiltonian $H_{\rm S-RC}$
\begin{align}
    \dot{x}_{\rm S} &= p_{\rm S},\\
    \dot{p}_{\rm S} &= -\Omega_1^2 x_{\rm S} + g X_{\rm R}, \\
    \dot{X}_{\rm R} &= P_{\rm R},\\
    \dot{P}_{\rm R} &= g x_{\rm S} - \Omega_2^2 X_{\rm R}.
\end{align}

\subsection{The GKLS Master Equation for the Bare System}\label{app:localGKLSME}
We now formulate the standard local GKLS master equation for the original system-bath Hamiltonian given in Eq.~\eqref{eq:original_ham_appendix}. As discussed in the main text, this standard approach fails in the strong-coupling regime, which necessitates the use of the global RC master equation derived in Appendix \ref{app:GKLSME}.

To solve the open-system dynamics, we again employ the Gaussian maps in Eq.~\eqref{eq:Gaussian_map}, which are fully determined by evaluating the drift matrix $A$ and the diffusion matrix $D$. We define the standard phenomenological Lindblad jump operators as \cite{PhysRevA.94.052129, wiseman_milburn_2009}
\begin{align}
    L_1 &= \left[ \kappa (\mathcal{N} (\Omega_{\rm S}, T) +1) \right]^{1/2} a, \\
    L_2 &= \left[ \kappa \mathcal{N} (\Omega_{\rm S}, T) \right]^{1/2} a^\dagger,
\end{align}
where the decay rate is given by $\kappa \coloneqq \mathcal{J} (\Omega_{\rm S}) / \Omega_{\rm S}$. The corresponding coefficient matrix $C$ is constructed as
\begin{equation}
    C = \begin{pmatrix}
        2^{-1/2} \left[ \kappa \Omega_{\rm S} (\mathcal{N} (\Omega_{\rm S}, T)+1) \right]^{1/2} & 2^{-1/2} \left[ \kappa \Omega_{\rm S} \mathcal{N} (\Omega_{\rm S}, T) \right]^{1/2} \\[1.5ex]
        i (2 \Omega_{\rm S})^{-1/2} \left[ \kappa \mathcal{N} (\Omega_{\rm S}, T) \right]^{1/2} & -i (2 \Omega_{\rm S})^{-1/2} \left[ \kappa \mathcal{N} (\Omega_{\rm S}, T) \right]^{1/2}
    \end{pmatrix}.
\end{equation}
Using this matrix, the drift and diffusion matrices evaluate to
\begin{equation}
    A = \begin{pmatrix}
        - \kappa / 2 & 1 \\[1.5ex]
        - \Omega_{\rm S}^2 & - \kappa / 2
    \end{pmatrix}, \quad D = \begin{pmatrix}
        \kappa \Omega_{\rm S}^{-1} \left( \mathcal{N} (\Omega_{\rm S}, T) + 1/2 \right) & 0 \\[1.5ex]
        0 & \kappa \Omega_{\rm S} \left( \mathcal{N} (\Omega_{\rm S}, T) + 1/2 \right)
    \end{pmatrix}.
\end{equation}
Given these explicit matrices, we fully simulate the local dynamics of the system via the time evolution of its first and second moments according to Eq.~\eqref{eq:Gaussian_map}.

\subsection{Comparison of Exact and RC-Mapped Dynamics}\label{app:comparisondynamics}
We now numerically evaluate the exact covariances from Eqs.~\eqref{eqa11} and \eqref{eqpS} and compare them against both the global RC master equation (Appendix \ref{app:GKLSME}) and the local bare master equation (Appendix \ref{app:localGKLSME}). This comparison explicitly demonstrates the superiority of the reaction coordinate mapping in the strong-coupling regime.

\begin{figure*}[h]
    \centering
        \includegraphics[width=.49\linewidth]{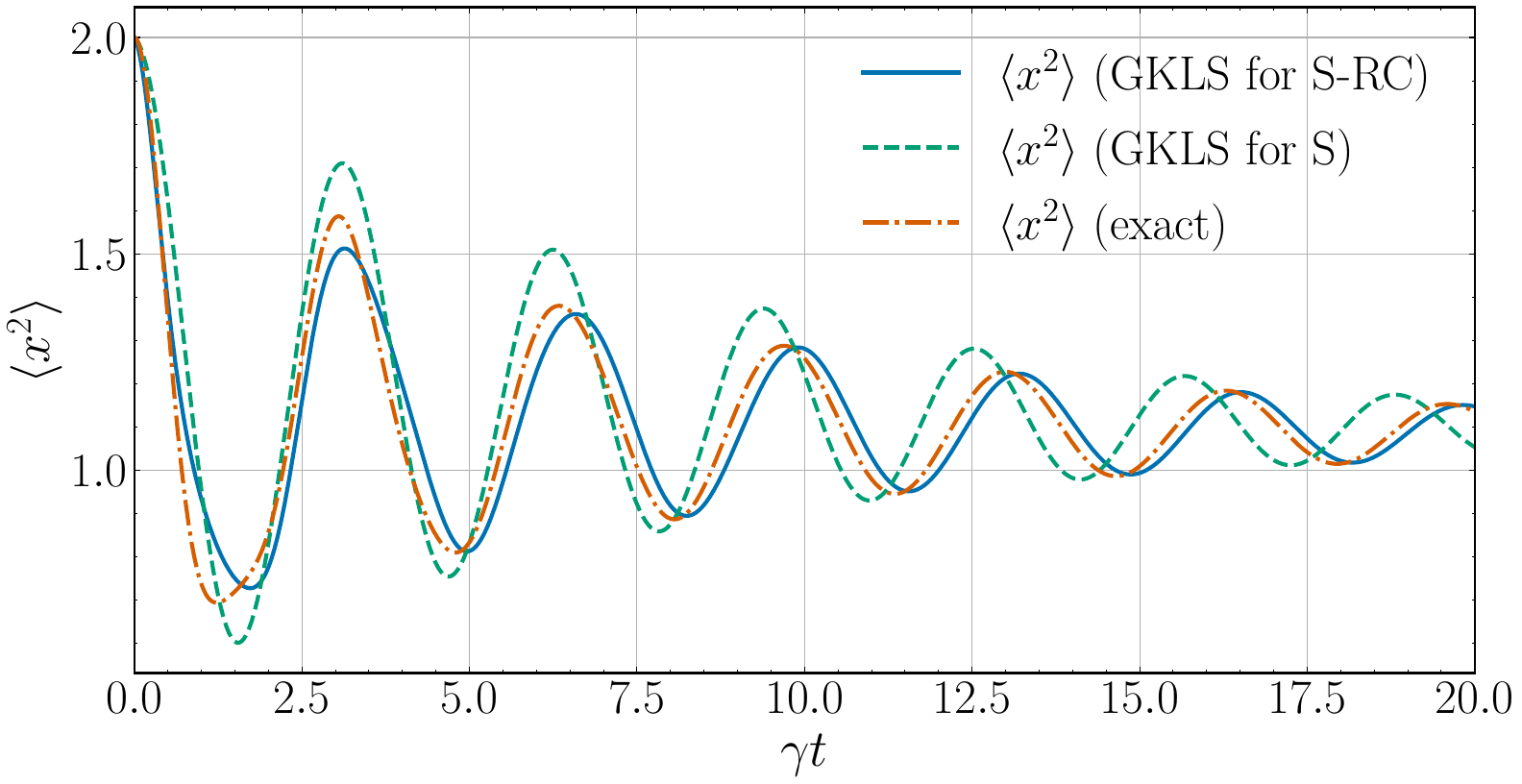}
        \includegraphics[width=.49\linewidth]{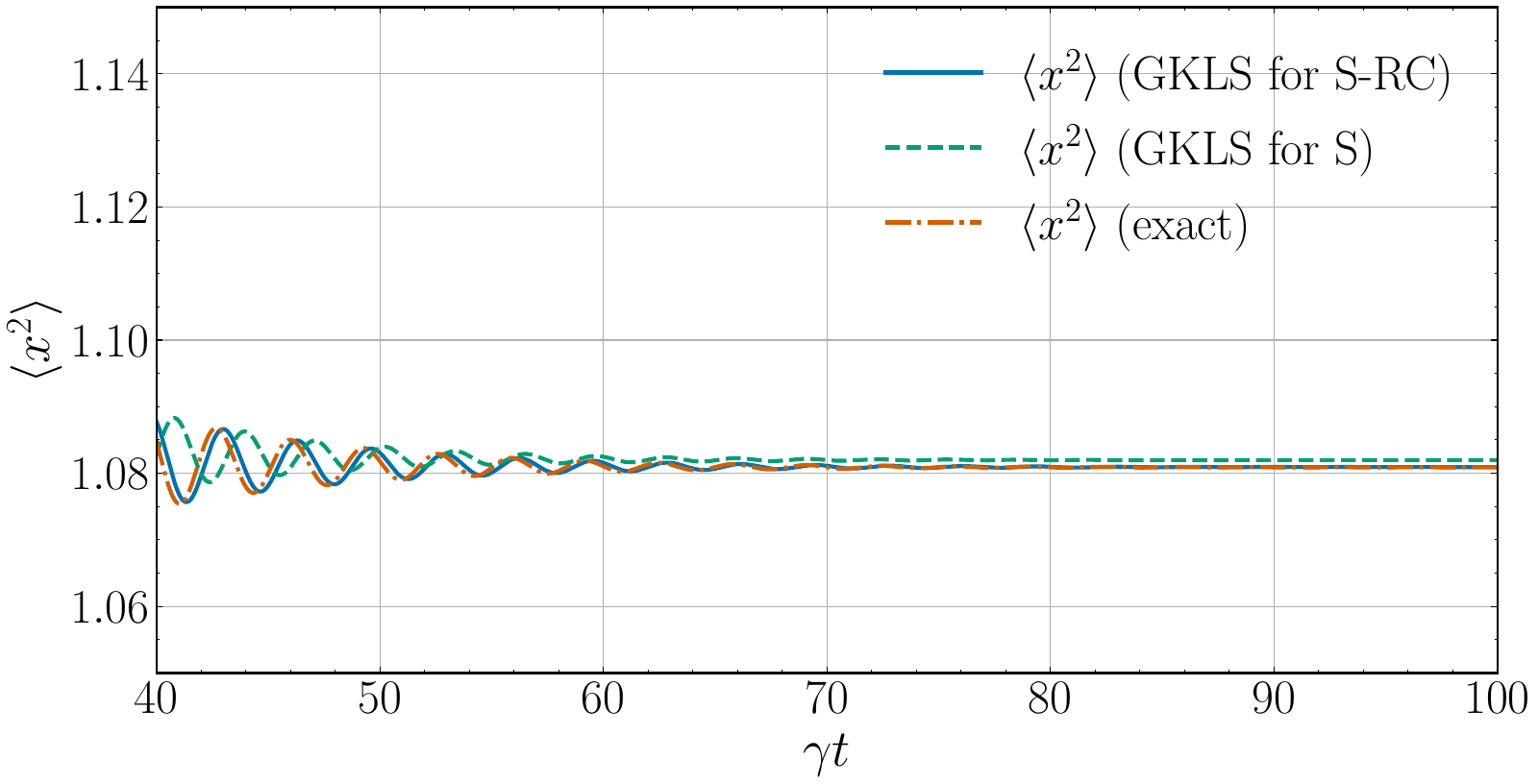}
        \includegraphics[width=.49\linewidth]{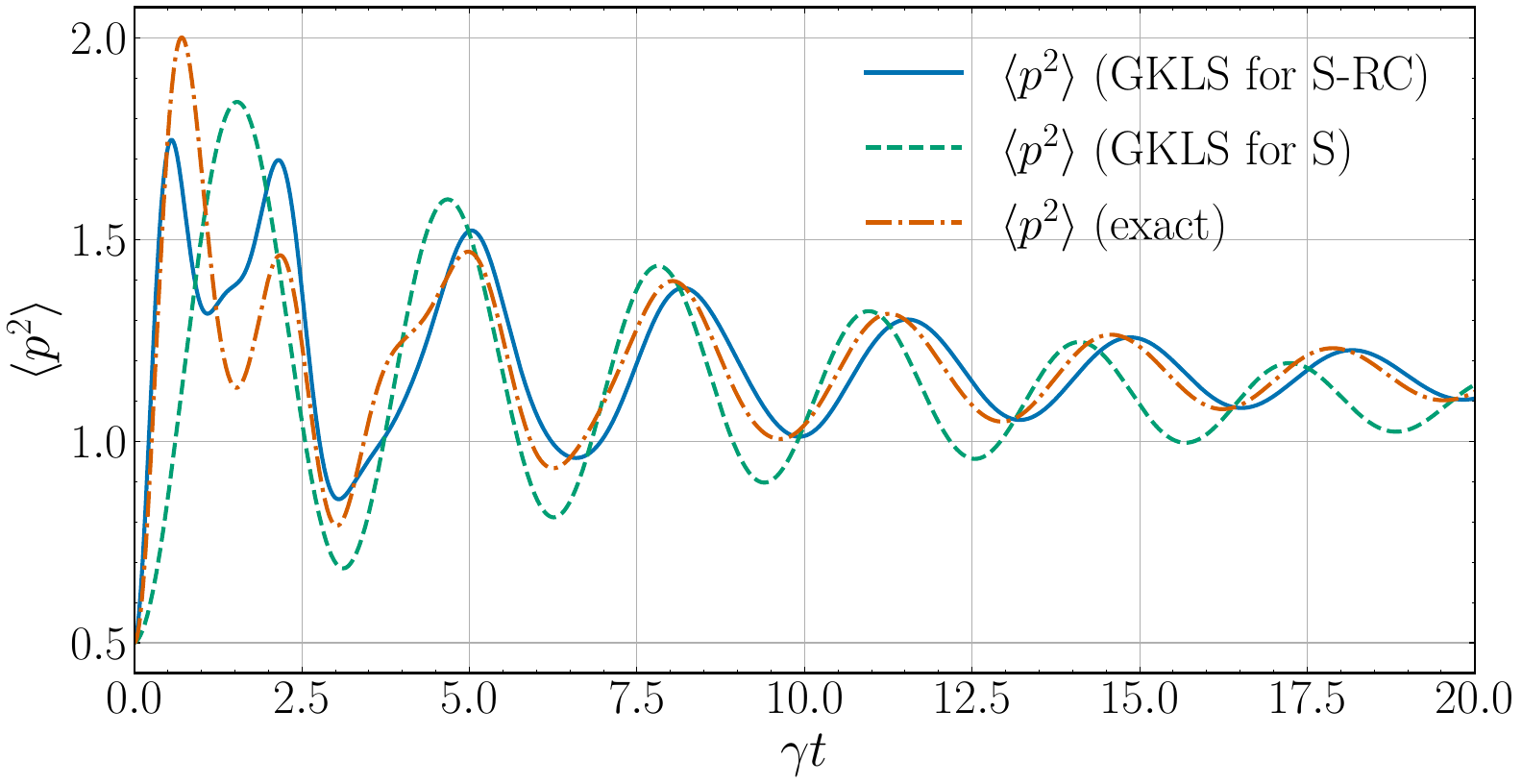}
        \includegraphics[width=.49\linewidth]{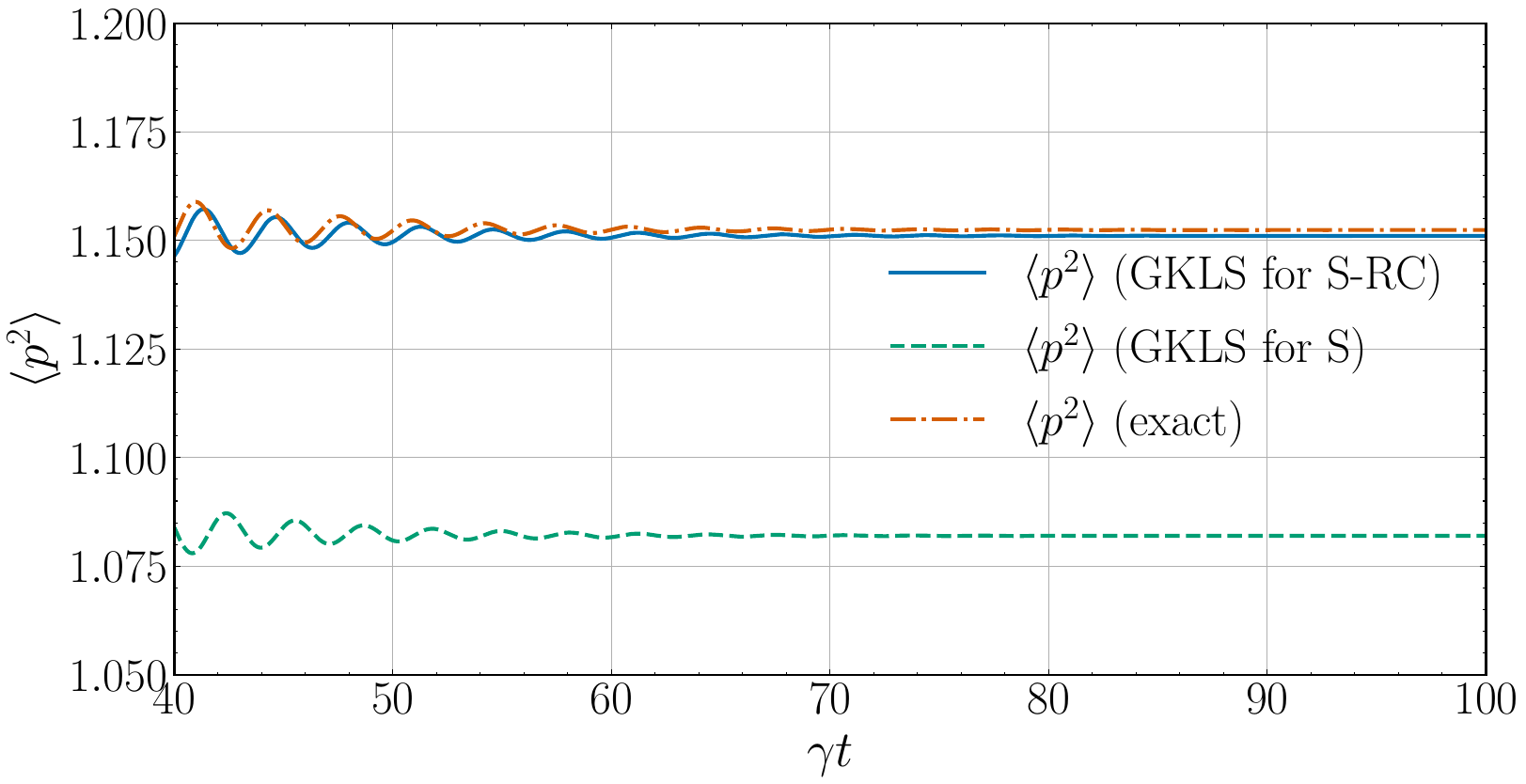}
    \caption{Comparison of the position and momentum variances evaluated using the exact quantum Langevin equation, the global RC master equation, and the local bare master equation. System parameters are set to $\Omega_{\rm S} = 1.0, \alpha_1 = 1, \alpha_2 = 10, \gamma = 1, \text{and } T = 1$.}
    \label{fig:variance_comparison_both}
\end{figure*}

In Fig.~\ref{fig:variance_comparison_both}, we plot the dynamical evolution of the system variances for all three approaches, capturing both the early-time transient regime and the long-time steady state. As expected, the global RC master equation successfully tracks the relevant transient oscillatory features and closely approaches the exact steady-state values. Conversely, the local approach fails to fully capture the exact dynamics, as it cannot account for strong system-bath correlations. We choose $\sigma_0^{\rm S} = \rm{diag}(2/\Omega_{\rm S}, \Omega_{\rm S}/2)$ as the initial system covariance matrix, both for the local GKLS and exact solutions. For the global GKLS master equation, we use $\sigma_0^{\rm S-RC} = \rm{diag} (2/\Omega_{\rm S}, \Omega_{\rm S}/2, 2/\Omega_{\rm R}, \Omega_{\rm R}/2)$. 




\end{document}